%% file: mobility_edge.tex
\documentclass[prb,twocolumn,english,superscriptaddress,footinbib,floatfix]{revtex4-2}

\usepackage{amsmath,amssymb,mathrsfs}
\usepackage{amsfonts,bm}
\usepackage{amsthm}
\usepackage{graphicx}
\usepackage{graphics}
\usepackage{subfigure}
\usepackage{color}
\usepackage{bm}
\usepackage{hyperref}
\usepackage{rotating}
\usepackage{comment}

\usepackage{xfrac}

\usepackage{diagbox}
\usepackage{slashbox}
\usepackage{mathtools}
\usepackage{bbm}
\usepackage{enumerate}
\usepackage{tikz}
\usetikzlibrary{external}
\usepackage[all]{xy}

\usepackage{etoolbox}

\usepackage{xr}

\usepackage{subfiles}

\newcommand{\bra}[1]{\langle #1|}
\newcommand{\ket}[1]{|#1\rangle}
\newcommand{\braket}[2]{\langle #1|#2\rangle}

\newcommand{\be}{\begin{equation} }
\newcommand{\ee}{\end{equation} }
\newcommand{\ba}{\begin{eqnarray} }
\newcommand{\ea}{\end{eqnarray} }

\newcommand{\bpm}{\begin{pmatrix}}
\newcommand{\epm}{\end{pmatrix}}
\newcommand{\bmm}{\begin{matrix}}
\newcommand{\emm}{\end{matrix}}

\newcommand{\bea}{\begin{eqnarray}}
\newcommand{\eea}{\end{eqnarray}}

\makeatother

\begin{document}
\newcommand{\op}[1]{\operatorname{#1}}

\title{In search of a many-body mobility edge with matrix product states in a generalized Aubry-Andr\'e model with interactions}

\author{Nicholas Pomata}
 \email{pomata@umd.edu}
 \altaffiliation[Current affiliation: ]{Joint Quantum Institute, University of Maryland at College Park, MD US 20742}
 \affiliation{C. N. Yang Institute for Theoretical Physics and Department of Physics and Astronomy, State University of New York at Stony Brook, NY 11794-3840, United States}  
\author{Sriram Ganeshan}
 \affiliation{Physics Department, City College of New York, NY 10031}
 \affiliation{CUNY Graduate Center, New York NY, 10031}

\author{Tzu-Chieh Wei}
 \affiliation{C. N. Yang Institute for Theoretical Physics and Department of Physics and Astronomy, State University of New York at Stony Brook, NY 11794-3840, United States}  

\vfill
\begin{abstract}

We investigate the possibility of a many-body mobility edge in the  generalized
Aubry-Andr\'e (GAA) model with interactions using the
Shift-Invert Matrix Product States (SIMPS) algorithm [Phys. Rev. Lett. 118,
017201 (2017)]. The non-interacting GAA model is a one-dimensional
quasiperiodic model with a self-duality-induced mobility edge. To search for a
many-body mobility edge in the interacting case, we exploit the advantages of
SIMPS that it targets many-body states in an energy-resolved fashion and
does not require all many-body states to be localized for some to converge.
Our analysis indicates that the targeted states in the presence of the
single-particle mobility edge match neither `MBL-like' fully-converged
localized states nor the fully delocalized case where SIMPS fails to converge.
We benchmark the algorithm's output both for parameters that give fully converged,
`MBL-like' localized states and for delocalized parameters where SIMPS fails to
converge. In the intermediate cases, where the parameters produce a single-particle mobility
edge, we find many-body states that develop entropy oscillations as a function of cut
position at larger bond dimensions.
These oscillations at larger bond dimensions, which are also found in the fully-localized benchmark but not the
fully-delocalized benchmark, occur both at the band edge and center and may indicate
convergence to a non-thermal state (either localized or critical).

\end{abstract}

\maketitle


\section{Introduction}

Isolated quantum systems are conjectured to equilibrate at the level of a
single eigenstate via subsystem thermalization in the absence of a bath. This
conjecture is known as the Eigenstate Thermalization Hypothesis
(ETH)~\cite{deutsch1991quantum, srednicki1994chaos}. Over the past decade,
many-body localization (MBL) has emerged as a candidate phase that  maximally
violates ETH, where all the eigenstates fail to equilibrate at the subsystem
level.  Many agree that MBL exists in one dimension with
short-range interactions ~\cite{basko2006metal, imbrie2014many,
imbrie2016diagonalization}, and experiments indicate the existence of MBL in a
number of platforms~\cite{schreiber2015observation, bordia2016coupling,
chiaro2019growth}. However, a recent challenge poses that the
localization effects seen in exact-diagonalization studies may result from
finite-size effects which will be destroyed by quantum chaos at sufficiently
large length scales~\cite{weiner2019slow,suntajs2020chaos,
schulz2020phenomenology, taylor2020subdiffusion, abanin2021distinguishing, sels2021dynamical,
sels2021markovian, morningstar2022avalanches, sierant2022challenges}, and
how to unambiguously quantify MBL in an experiment
is still a work in progress.   

A natural question then arises as to whether MBL always represents the most
generic violation to ETH, where all eigenstates are non-thermal, or there can
be cases where only part of the many-body spectrum will be localized.  Exceptions to this case have been found in the form of quantum many-body scar states where a sub-extensive number of area law entangled states~\cite{turner2018quantum} interspersed among an extensive number of volume law states. 
A full many-body mobility edge with extensive localized and delocalized states separated by critical energy  was originally presented in the works of
Basko, Aleiner and Altshuler~\cite{basko2006metal} where they found a possible
many-body delocalization phase transition at finite temperature. Numerical
works have observed evidence for a many-body mobility edge~\cite{bardarson2014,
li2015many, modak2015many, serbyn2015criterion, luitz2015many,
geraedts2017emergent, goihl2019exploration, yao2020many, brighi2020stability,
PhysRevResearch.2.032045},
although finite size effects plague the reliability of these results. However,
the works of De Roeck et al. \cite{mobility-absence} claim to exclude the
possibility of \textit{any} mobility edge using avalanche arguments.   More
recently, experiments have shown evidence for a many-body mobility edge in a
shallow lattice limit of the Aubre-Andr\'e model~\cite{luschen2018single,
kohlert2019observation}. It is an open question if the experimental observation
of a non-ergodic phase is an indication of a more robust violation of ETH or
simply a finite-size and finite-time effect. The question of the presence
or absence of many-body mobility edges remains unresolved, although
the experimental capability of energy resolution can potentially offer further
advancement~\cite{guo2021observation}.

In this paper, we investigate the fate of many-body localization in the presence
of a single particle mobility edge at large system sizes. We consider the
interacting version of the generalized Aubry-Andr\'e (GAA) model of Ref.~\cite{GAA1}, which
possesses a mobility edge protected by self-duality in its single particle
spectrum. Machine learning methods have indicated the existence of a non-ergodic
metal in the center of the many-body spectrum of this
model~\cite{hsu2018machine}. Recent experiments have realized the bosonic
version of the GAA model in the synthetic lattices of laser-coupled atomic
momentum modes and studied the influence of weak interactions on the mobility
edge~\cite{an2020observation}. In order to address this question at large
system sizes, we use the energy-targeting Shift-Invert Matrix Product States
(SIMPS) algorithm of Yu, Pekker, and Clark~\cite{SIMPS}. We show that the SIMPS
method should be capable of identifying a many-body mobility edge due to its
energy-targeting nature. We benchmark the properties of the targeted matrix
product state (MPS) in the mobility-edge regime to that of the convergent
fully-localized regime and the fully-delocalized regime where the algorithm is
expected to fail to converge the delocalized energy eigenstates. 

We find that the single-cut entanglement entropy shows oscillations in the cut
location which appear at higher bond dimensions. Similar oscillations are
typically seen in critical (logarithmic entanglement
scaling)~\cite{laflorencie2006boundary} states with open boundary conditions.
This phenomenon is not observed in the fully delocalized case as benchmarked
with SIMPS within the bond dimensions considered. Where observed, entanglement
oscillations are stronger for the states at the band edge but are also present
near the band center and may indicate convergence to some kind of non-thermal
state whose exact nature is difficult to quantify within our methods.

The remainder of the paper is structured as follows. In Sec.~\ref{sec:GAA}, we introduce the 
generalized Aubry-Andr\'e (GAA) model
study in this paper, whose non-interacting version exhibits single-particle
mobility edge protected by self-duality.
In Sec.~\ref{sec:SIMPS}, we briefly describe the SIMPS method (relegating a detailed account of the numerical procedure  to Appendix~\ref{app:algorithm}) and benchmark it for two cases with clearly
localized and thermalized behavior, respectively.
Then in Sec.~\ref{sec:candidates}, we compare these benchmarks to candidate eigenstates produced
by SIMPS in the neighborhood of the single-particle mobility edge and analyze the average entanglement scaling.  We conclude in Sec.~\ref{sec:Conclusions}.
In addition to the details of the algorithm, the Appendix contains the analysis of additional datasets
(including ones with greater system size); details of 
the calculation of the energy error and single-cut entanglement entropy; and the analysis of
additional characterization by the Uhlmann fidelity.
\begin{figure}
  \includegraphics[width=.55\textwidth]{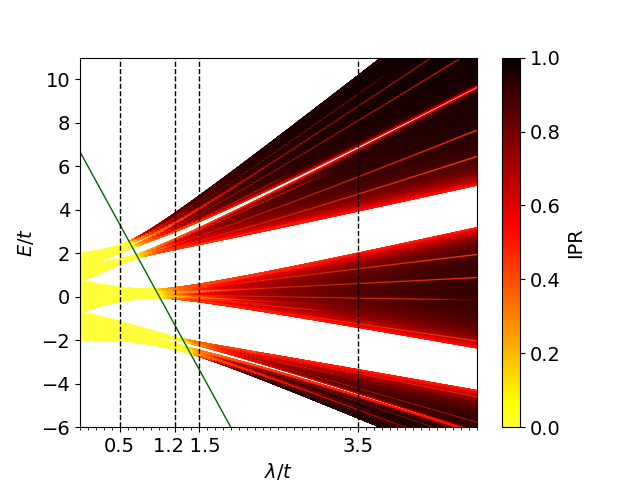}
  \caption{The spectrum and inverse participation ratio (IPR)
  of the noninteracting generalized Aubry-Andr\'e model at $\alpha=0.3$ with various
  $\lambda$ and any $t>0$, obtained by diagonalizing single-particle
  Hamiltonians. For the length, we use a Fibonacci number ${N=F_{16}=987}$, 
  which allows us to use periodic boundary conditions and minimize boundary
  effects with the frequency ${b=F_{15}/F_{16}}$ approximating the inverse
  golden ratio. The distinction between
  delocalized ($\text{IPR}\sim 1/N$) and localized ($\text{IPR} > 1/N$,
  increasing to 1) behavior on either side of the self-dual line $\alpha E =
  2(t-\lambda)$ (green) is clear. Disorder strengths $\lambda$ studied herein
  in the interacting case are marked in black.}
  \label{fig:noninteracting}
\end{figure}

\section{The generalized Aubry-Andr\'e model}
\label{sec:GAA}
\subsection{The non-interacting case}
The generalized Aubry-Andr\'e model (GAA)~\cite{GAA1} we consider is defined, in the
non-interacting case, by the Hamiltonian
\begin{equation}
\begin{split}
H_0 = &\ t\sum_{n=1}^{L-1}(\psi_n^\dagger\psi_{n+1} + \psi_{n+1}^\dagger\psi_n)\\
&+2\lambda\sum_{n=1}^L\frac{\cos(2\pi b n+\phi)}{1-\alpha\cos(2\pi b n+\phi)}\psi_n^\dagger\psi_n,
\end{split}
\label{eq:GAA-ham}
\end{equation}
which becomes the standard Aubry-Andr\'e model when $\alpha=0$, with the phase
$\phi$ determining a family of ``disorder realizations''. We also use the
standard choice of $b$ as the inverse golden ratio $\frac{2}{1+\sqrt{5}}$. When
$t>0$ and $|\alpha|<1$, this has been determined to be self-dual for energies
\begin{equation}
\alpha E = 2(t-\lambda).
\end{equation}
One may diagnose localization in this model on either side of this self-dual line, e.g.
by using the inverse participation ratio (IPR), as shown in Fig.~\ref{fig:noninteracting}.
For a single-particle state $\ket{\psi}$ with wavefunction $\psi_i$, the inverse participation ratio is defined as
\begin{equation}
\mathrm{IPR} = \frac{\sum\limits_{i=1}^L|\psi_i|^4}{\left(\sum\limits_{i=1}^L|\psi_i|^2\right)^2}. 
\label{eq:ipr}
\end{equation}
When excitations are localized (to a region that does not scale with system
size), $\mathrm{IPR} \sim O(1)$, whereas thermalization implies
$\mathrm{IPR} \sim O(1/L)$. As predicted by self-duality, there is a
mobility edge for nonzero $\alpha$ at $E = \frac{2}{\alpha}(t-\lambda)$.

\subsection{The interacting model}
Later works considering an interacting version of this model
\cite{GAA2PRL,GAA2PRB,GAA2AdP}, constructed with the simple addition of a
four-fermion term
\begin{equation}
  H = H_0 + V\sum_n\psi_n^\dagger\psi_n\psi_{n+1}^\dagger\psi_{n+1},
  \label{eq:GAA-interaction}
\end{equation}
have analyzed it with exact diagonalization for small sizes and low filling factors. 

The main goal of this work is to expand the system size substantially using the
SIMPS method~\cite{SIMPS}. The tradeoff for large system sizes is that the
finite bond dimension cuts off the entanglement of the state. If the many-body
state obtained by SIMPS is localized then the entanglement does not scale
with the cut size and is unaffected by the increasing bond dimension. However,
a thermalized state should have volume-law entanglement scaling: in particular, the
half-cut entanglement entropy should be asymptotically proportional to the
system size. Since the bond dimension of an MPS is exactly the Schmidt rank across
a given cut, the single-cut entanglement entropy of an MPS with (maximal) bond dimension
$\chi$ is limited to precisely $\log\chi$, making the bond dimension of an adequate
MPS representation exponential in the system size as a function of the cut size.
Thus we trade system-size limitations suffered in exact diagonalization methods with
finite-entanglement limitations due to limited bond dimension. The advantage of
this tradeoff is that we can benchmark states against fully localized and fully
delocalized systems in terms of how their properties scale with the bond
dimension while making finite-size effects negligible.

\subsection{Model parameters}

\begin{figure}
  \includegraphics[width=.37\textwidth]{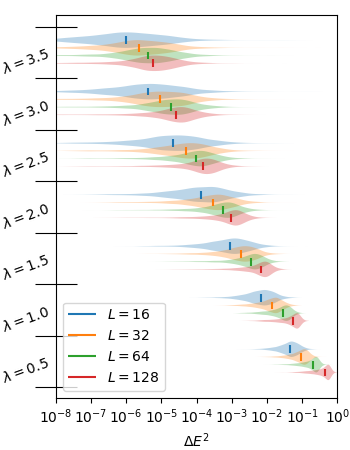}\\
  \caption{
  A violin plot displaying the distribution of energy
  errors at several system sizes and disorder strengths $\lambda$, given bond
  dimension ${\chi=10}$; tests sample the full range of the energy spectrum.
  One can tell that the distributions are \textit{qualitatively}
  different among the lowest three system sizes, whereas they are qualitatively
  similar for $L=64$ and $L=128$. (Meanwhile, a \textit{quantitative} comparison
  is complicated by a need to account for both extrinsic scaling of $E$ and
  exponential suppression of level spacing.)
  }
  \label{fig:bwchi10}
\end{figure}

\begin{figure}[t!]
\centering
  \includegraphics[width=.45\textwidth]{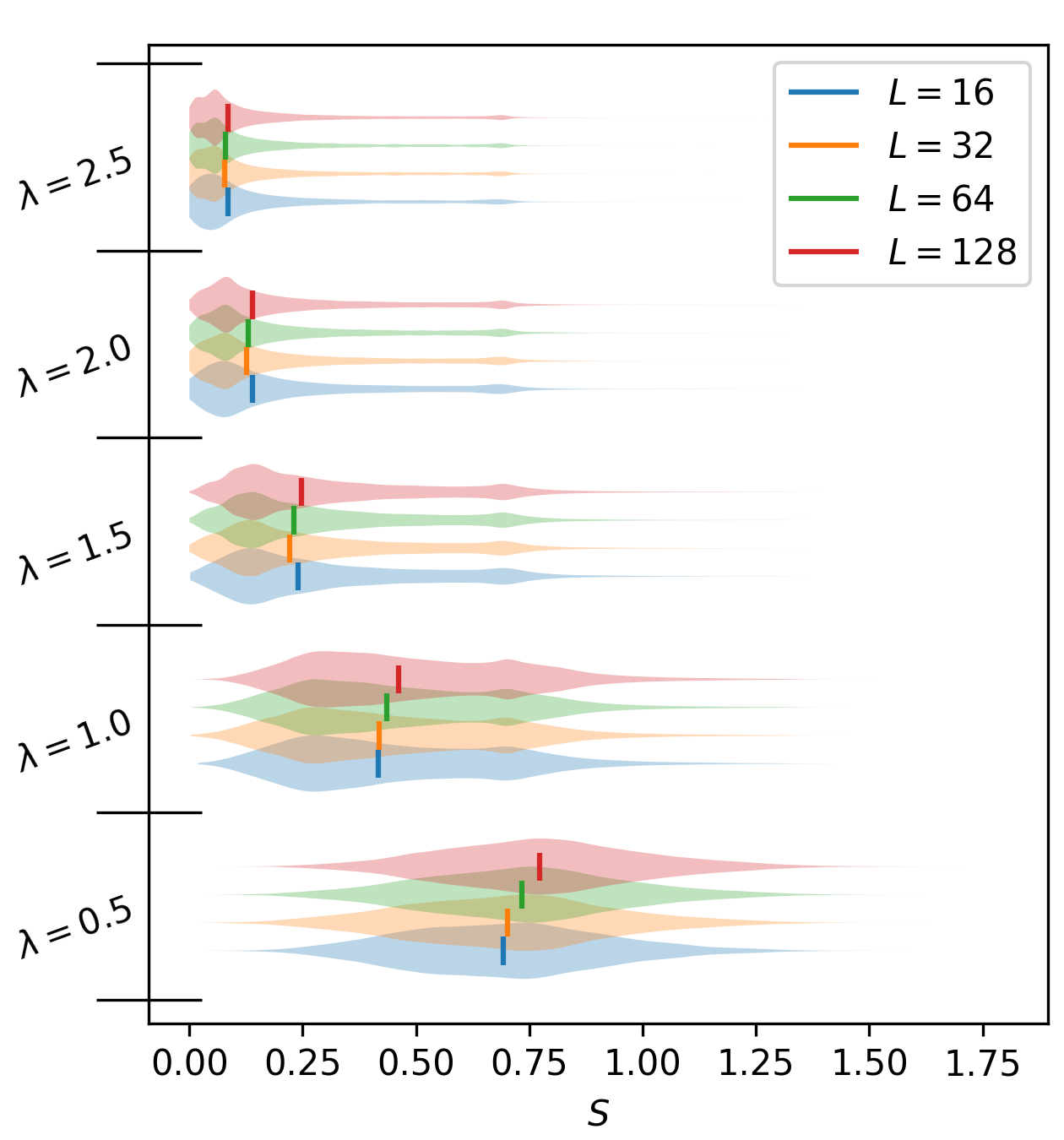}
  \caption{Entropy distribution for several values of the disorder strength $\lambda$, given various
  system sizes with bond dimension $\chi=10$. We find that, for a given $\lambda$, the behavior at different system sizes has few qualitative differences; we may find some for the thermalizing regime (here $\lambda=0.5$ and $\lambda=1$) between $L=16$ and $L=32$ and between $L=32$ and $L=64$, but only a modest and expected difference in the location of the median between $L=64$ and $L=128$. This helps us demonstrate that finite-size effects are minimal (particularly in comparison to finite-entanglement effects) by the time we reach $L = 64$.
  }
  \label{fig:Shistold}
\end{figure}

The systems we will be primarily considering will have an interaction strength
$V/t=1$ and a mobility-edge parameter $\alpha=0.3$. We enforce particle-number
conservation as a global $U(1)$ symmetry in order to restrict to half-filling.
In preliminary studies we have varied the system size: with half-integer disorder strengths
$\lambda=\{0.5,1,1.5,2,2.5,3,3.5\}$, six disorder
realizations $\phi=\pi n/3$, and system sizes $L=\{16,32,64,128\}$, we used
bond-dimension $\chi=10$ SIMPS to probe the system
at equally-spaced test energies encompassing the entire spectrum.
As shown in Figures~\ref{fig:bwchi10} and~\ref{fig:Shistold},
these studies have demonstrated to
our satisfaction that finite-size effects are sufficiently small for system sizes of order
$L \geq 64$. In  Appendix~\ref{app:sys-size}, we additionally compare data 
obtained from the primary studies (described below)---in particular those with smaller
bond dimensions---with tests run using the same parameters but larger bond dimensions, 
and determine that we do not see a significant qualitative difference.
Finally, we note that even the longest-lived boundary effects seen in this work, as found e.g. in Sec~\ref{sec:benchmarkii}
do not penetrate beyond a distance from the boundary of $\ell\sim 15$.
Thus, for the primary studies discussed herein, we select a fixed size $L=64$.

We choose sample ``disorder
strengths'' $\lambda$ of $0.5,1.2,1.5,3.5$. As illustrated in Fig.~\ref{fig:noninteracting}, 
these are, respectively, well before,
intersecting, just beyond, and well beyond the single-particle mobility edge.
The intention behind these choices is as follows:
\begin{itemize}
    \item $\lambda=0.5$ should be well within the thermalizing phase and therefore
    should provide a benchmark for SIMPS output in this paradigm (i.e. when eigenstates
    are volume-law and therefore unrepresentable as MPS).
    \item $\lambda=3.5$ should be well within the localized phase and therefore should
    provide a benchmark for SIMPS output in this paradigm (when eigenstates should be
    easily representable with MPS).
    \item $\lambda=1.2$ and $\lambda=1.5$, meanwhile, provide candidates for a mobility
    edge, wherein results can be compared to the above benchmarks to determine whether a
    given energy range corresponds to the localized or thermalized phase.
\end{itemize}

Finally, we select 12 disorder
realizations via phases $\phi=\pi n/6$. In each system, for each of the bond
dimensions 10,~14,~20,~25,~and~30, we sample 99 target energies equally spaced
within each of two energy ranges, determined as follows. We can readily approximate the
minimum and maximum energies $E_\mathrm{min}$ and $E_\mathrm{max}$ given half filling and fixed
$t,\lambda,V,\alpha,b,\phi,$ and $L$, $E_\mathrm{min}$ being the
ground-state energy of $H$ and $-E_\mathrm{max}$ being the ground-state energy of $-H$.
Then the energy densities $E_\mathrm{min}/L$ and $E_\mathrm{max}/L$ will have minimal
dependence on $\phi$ and $L$, so, fixing $t,V,\alpha,b$, we can define an
energy density above the ground state
\begin{equation}
    \varepsilon\equiv \frac{E-E_\mathrm{min}}{L}
    \label{eq:energydensitydef}
\end{equation}
which specifies $E$ for a given $L$ and $\lambda$ and which has 
\[ 0\leq \varepsilon\leq \varepsilon_\mathrm{max}\equiv (E_\mathrm{max}-E_\mathrm{min})/L.\]

We use ``lower'' and ``middle'' energy ranges
\begin{itemize}
    \item 
    $0.1\varepsilon_\mathrm{max}<\varepsilon<0.15\varepsilon_\mathrm{max}$
    (a target energy density of $\varepsilon_m=(0.1+0.0005m)\varepsilon_\mathrm{max}$ for $m=1,2,\ldots,99$)
    \item 
    $0.45\varepsilon_\mathrm{max}<\varepsilon<0.5\varepsilon_\mathrm{max}$
    (a target energy density of $\varepsilon_m=(0.4+0.0005m)\varepsilon_\mathrm{max}$ for $m=1,2,\ldots,99$)
\end{itemize}
Note that, for each energy range and each
value used of the ``disorder'' strength
$\lambda$ and the bond dimension $\chi$, we have a sample size of 1188 states.

\section{The numerical method}
\label{sec:SIMPS}
Friesdorf et al.~\cite{mps-proof} have shown that matrix product states
can efficiently represent excited eigenstates of localized systems and are
therefore an effective means of non-perturbatively analyzing localized systems
at large system sizes. To extract MPS approximations of eigenstates, we use the
SIMPS algorithm~\cite{SIMPS}, which we outline in detail in Appendix~\ref{app:algorithm}.
SIMPS and other MPS algorithms can only attempt
to diagonalize a system under the assumption that it is localized, meaning that
they will otherwise give ``false positives'' of relatively low-entanglement
states which are not approximations of any eigenstates and are instead linear
combinations of states with similar entropies.
{
Indeed, if we assume a typical energy spacing, at a typical energy $E$, of
$s\sim 2^{-L}E$ (where $L$ is the
system size), an equal combination of $n$ adjacent states would have energy
variance on the order of $\Delta E^2/E^2 \sim \frac{n^2}{12}2^{-2L}$ (see
\eqref{eq:energy-error-calc} of the Appendix). At a system size
$L=64$, such a combination could still have energy error below machine
precision if it consisted of $2^{33}$ distinct eigenstates. However, our
intuition and the benchmarks we use suggest this is not the case, e.g., a
low-energy-error superposition like that would still have comparably high entropy and
thus could not be replicated as an MPS. }

\begin{figure}
  \includegraphics[width=.5\textwidth]{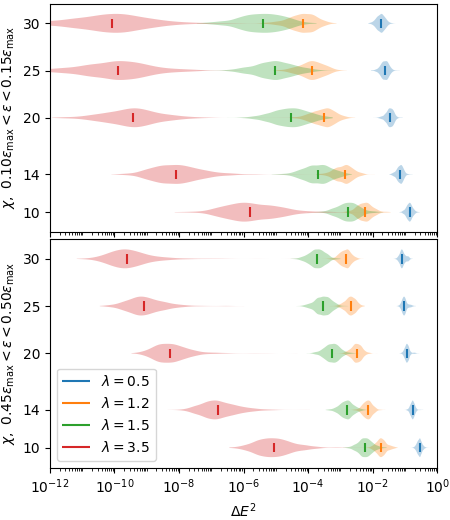}
  \caption{A violin plot representing the distribution of energy error for
  various disorder strengths; the area of a shape in a given region
  approximates the frequency of samples within the 1188 sample states, and
  lines are placed at the medians. In all cases, the error substantially
  decreases with bond dimension; however, the errors at $\lambda=0.5$ and
  $\lambda=3.5$ are, for all bond dimensions, at substantially different
  scales. The energy errors for $\lambda=1.2$ and $\lambda=1.5$, meanwhile,
  only approach either scale at high bond dimension, when the $\lambda=1.5$
  low-energy case begins to overlap low-bond-dimension results from the
  localized case.}
  \label{fig:energy-violin}
\end{figure}

\begin{figure}
  \includegraphics[width=.5\textwidth]{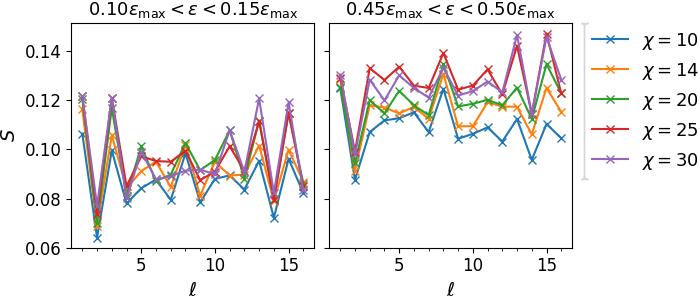}
  \caption{Single-cut entropy by cut location (i.e. distance from the nearest
  endpoint) for strong disorder, $\lambda=3.5$, at $L=64$. The entropy has
  evidently saturated by $\chi=30$ for the different cuts and energies, under
  consideration, and it
  remains low moving into the bulk, as is generally expected for the area-law
  behavior of localized states.}
  \label{fig:entropy-cut3.5}
\end{figure}

We note that, in the similar case of MPS approximations to critical ground
states, there exist well-established scaling
relations~\cite{critical-entanglement1,critical-entanglement2}. These relations
include the asymptotic behavior of the correlation length with respect to (a)
the single-cut entanglement entropy, (b) the bond dimension, and (c) the energy
error relative to the true ground state. Such a relation, applied to excited
states of disordered ergodic systems, would be necessary in order to
distinguish with any certainty the phases we hope to observe. In the absence of
such an asymptotic description, we attempt to extract empirical relationships
and benchmarks with respect to fully localized and fully delocalized cases
which can help separate localized and ergodic phases.

\subsection{Benchmark I: SIMPS applied to a fully-localized many-body system}
We begin by applying SIMPS to the disorder strength $\lambda=3.5$, that is, we
tune the system to be far into the region corresponding to single-particle
localization. In Fig.~\ref{fig:energy-violin}, we see that the states we find
given these parameters have very low energy variance $\Delta E^2$. In fact,
for high bond dimensions $\chi > 20$, $\Delta E^2$ appears to saturate at about
$10^{-10}$, of order comparable to the tolerance of the subroutines of our SIMPS
implementation. We then consider entanglement entropy, as in
Fig.~\ref{fig:entropy-cut3.5}. We see, first, that as a function of bond
dimension the entropy has also largely saturated by $\chi = 30$ (in fact,
in the entropy histograms discussed in Appendix~\ref{app:entropy} we find the entropy distribution has largely converged
with respect to bond dimensions). Indeed, the movement we see before that point
is likely attributable to a reduction of bias against higher-entropy states.
Moreover, we observe that neither the single-cut entropy nor the bond-dimension
corrections to it grow significantly as we move into the bulk of the system,
ruling out the possibility that our evidence of localization can be viewed
primarily as a finite-size effect.

\begin{figure}
  \includegraphics[width=.5\textwidth]{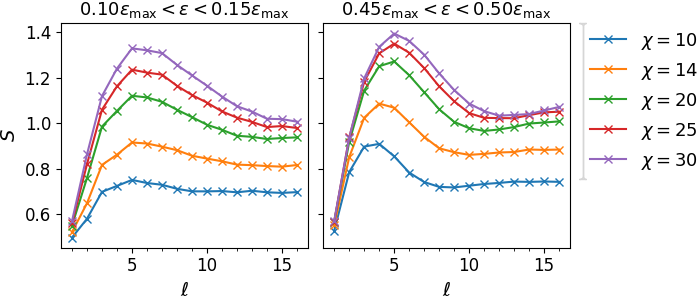}
  \caption{Single-cut entropy by cut location for weak disorder,
  $\lambda=0.5$, at $L=64$. While there are signs of convergence (to volume-law
  behavior) very close to the boundary, as seen in greater detail in
  Fig.~\ref{fig:trend-by-cut}, in general the entropy is high and failing to
  converge. Also notably, it peaks near the boundary, at a location $\ell\sim 5$
  that grows with the bond dimension, and then falls to a bulk value,
  suggesting a higher-order artifact of finite-entanglement scaling beyond
  simply limiting the entropy.}
  \label{fig:entropy-cut0.5}
\end{figure}

We also point out a distinct feature in the single-cut entropy displayed in Fig.~\ref{fig:entropy-cut3.5}, namely, the highly discrete, oscillatory behavior as the cut size $l$ varies. This feature is absent in the weak-disorder case, e.g., $\lambda=0.5$ in Fig.~\ref{fig:entropy-cut0.5}.

\subsection{Benchmark II: SIMPS applied to a fully delocalized system} 
\label{sec:benchmarkii}
The SIMPS algorithm naturally fails with any finite bond dimension for the
fully delocalized case due to volume law entanglement scaling. Nonetheless, we can
quantify this failure in the form of energy variance and entanglement scaling
with bond dimension. Within our model, the ``disorder'' strength $\lambda=0.5$
(and other parameters as above) corresponds to full delocalization in the
single-particle case. We find in Fig.~\ref{fig:energy-violin} for the system size $L=64$ that the energy
errors are very large, eclipsing the values that would be predicted by
na\"ively combining the cutoff error and density of states. The implication of
this is promising: even as the system size becomes large, the algorithm cannot
produce pseudo-eigenstates of the delocalized system which exploit tight energy
spacings to exhibit small energy error. 

In Fig.~\ref{fig:entropy-cut0.5}, we find rapid growth of entanglement entropy
as we move up to 5 sites into the bulk of the system; notably, while a failure
to converge is apparent away from the boundary, near the boundary we see
convergence to something resembling a volume law. The full entanglement distribution is given in Appendix~\ref{app:entropy}. Further away from the
boundary, we see the entanglement entropy fall again before settling into
asymptotic behavior; this is evidently an artifact of finite entanglement given
that the peak moves away from the edge as we increase the bond dimension.

We emphasize that in the weak-disorder regime studied in this section, the single-cut entropy vs. the cut size $l$ is smooth, in contrast to the previous larger disorder strength case with $\lambda=3.5$, where there were distinct, highly discrete oscillations. 
\begin{figure}
  \begin{subfigure}
    \centering
    \includegraphics[width=.5\textwidth]{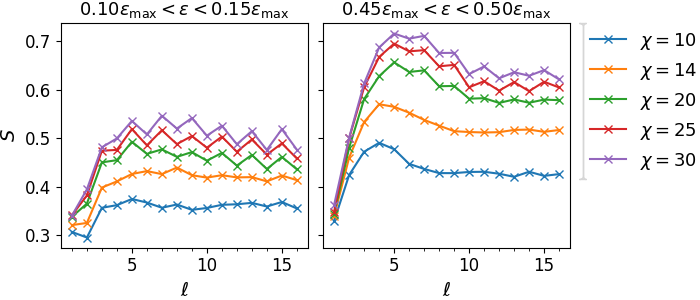}
  \end{subfigure}
  \begin{subfigure}
    \centering
    \includegraphics[width=.5\textwidth]{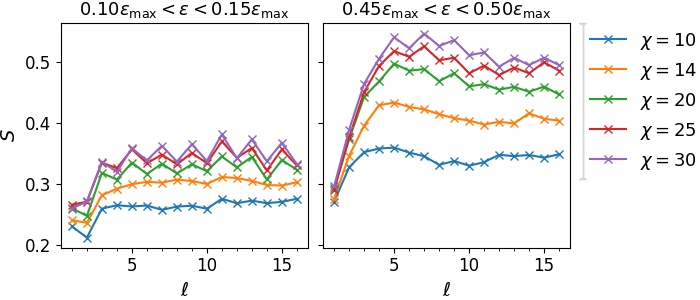}
  \end{subfigure}
  \caption{Single-cut entropy by cut location for disorder strengths
  $\lambda=1.2$ (top) and $\lambda=1.5$ (bottom) both at $L=64$. Convergence within the bulk
  by $\chi=30$ is apparent only in the lower energy range with $\lambda=1.5$.
  In the higher energy range, meanwhile, we observe for both $\lambda$
  convergence towards the boundary to apparently volume-law behavior as well as
  a clear peak in the entropy at $\ell\sim 5$, as is also seen with $\lambda=0.5$
  in Fig.~\ref{fig:entropy-cut0.5}.
  The substantial oscillations seen in the entropy here, including
  an observed fall-off between the first and second sites seen for smaller
  bond-dimension in the cases nearer the band edge, are discussed in more
  depth in Appendix~\ref{app:osc}.}
  \label{fig:entropy-intermediate}
\end{figure}

\section{Evaluating candidate disorder strengths for a mobility edge} 
\label{sec:candidates}
Now we present the main results of this paper. In the presence of the
single-particle mobility edge, the localization properties of the many-body
interacting states can in principle have four outcomes: (1) all
many-body states are localized; (2) the many-body spectrum has a mobility edge;
(3) all many-body states are delocalized;  and  (4) the spectrum contains
spectrum containing non-ergodic extended states (the exotic case). Even though the SIMPS
algorithm cannot unambiguously discriminate among all of these four scenarios,
it can locate the existence of localized states in the many-body spectrum in an
energy-resolved way. Thus we can address the question of whether the many-body spectrum contains any localized states when the single-particle spectrum
possesses a mobility edge within numerically accessible bond dimensions.  

We consider the disorder strengths $\lambda=1.2$, for which a full
single-particle band is delocalized, and $\lambda=1.5$, which is fully
localized (but with longer localization length in bands closer to the critical
line of the mobility edge than in benchmark I, $\lambda=3.5$), as shown in
Fig.~\ref{fig:noninteracting}.  In Fig.~\ref{fig:energy-violin} we have seen that the
energy error in either case does not truly match either the localizing or the
thermalizing case, and it is not clear from a qualitative analysis which
comparison is stronger.
We also see
evidence in Fig.~\ref{fig:entropy-intermediate} that the states in  middle of
the spectrum, where we see convergence toward volume-law entropy scaling
up to $\ell\simeq 4$, are more delocalized than those near the edge of the band,
where we only see a hint of entropy scaling with cut size.

\begin{figure}
  \includegraphics[width=.45\textwidth]{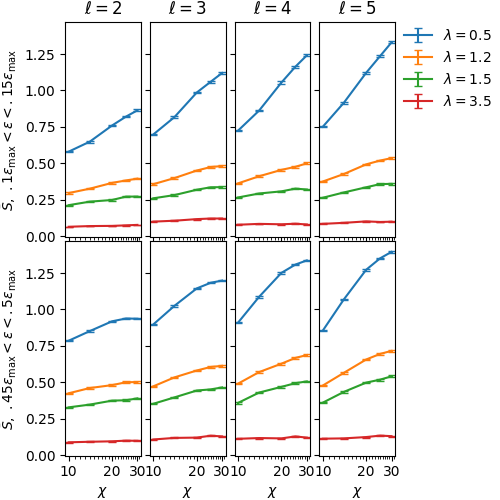}
  \caption{Average single-cut entanglement entropy, together with the standard error of the mean, as a function of the bond dimension $\chi$ for several
  small cuts in the various cases considered. We expect to see logarithmic
  growth with eventual convergence either to a constant value (in the
  localized, area-law case) or a value proportional to the distance from the
  boundary (in the delocalized, volume-law case). The asymptotic behavior
  generally anticipated in the bulk is logarithmic growth if the entropy is
  volume-law and convergence if it is area-law; near the boundary, in the
  volume-law case, we expect convergence to a value determined by that
  volume law.}
  \label{fig:trend-by-cut}
\end{figure}

In Fig.~\ref{fig:trend-by-cut}, we plot entanglement entropy versus bond
dimension at small cut sizes ($\ell=2,3,4,5$) averaged over the energy windows
selected near the band edge and center (the full distribution of entanglement entropy for
this case is explored in Appendix~\ref{app:entropy}). 
Within the bond dimensions we have used, we only observe
saturation of entanglement entropy with bond dimension (as occurs in benchmark I) in the
band-edge case of $\lambda=1.5$. In
the absence of such saturation, we can still use the dependence of
entanglement on both cut sizes (i.e., for single cuts, the distance between the
cut and the boundary) and bond dimension. 

For the case of $\lambda=1.2$, we see in
Fig.~\ref{fig:entropy-intermediate} that the average entropy at both the band
center and the band edge seem to contain features from both benchmark I (fully localized case) and II (the fully
delocalized case). However, with the increasing bond dimensions it seems to develop more features of the localized state. 
For small bond dimensions, the entanglement curves are somewhat smooth for both energy ranges we have probed. But as the bond dimension increases, the converged curves begin to exhibit highly discrete oscillation, similar to the regime of large disorder strength. Such oscillation in the entanglement entropy was observed previously in the ground state of the XXZ spin chain~\cite{laflorencie2006boundary} for open boundary conditions, and  identified as a dimerization process universal in ground states of models with a Luttinger
liquid description. We suspect that the presence of oscillations due to open boundary conditions may indicate the beginning of the convergence of the SIMPS algorithm toward capturing a faithful MPS representation. Note that the success of SIMPS at a finite bond dimension itself is evidence of the non-thermal nature of the state. On the contrary, for a thermal or a fully delocalized state, one would expect these oscillations to develop only at bond dimensions of the order $\chi\sim 2^{L/2}$.

\section{Conclusions}
\label{sec:Conclusions}
We have observed what appears to be a compelling distinction between
thermalized and localized behavior in an interacting quasiperiodic system at a
reasonably large system size $L=64$. In particular, we find that we can extract
``good'' eigenstates with low entropy when the strength of the quasiperiodic
``disorder'' is high; conversely, when it is low, we only find eigenstates of
poor quality (as measured by the energy error) whose entanglement entropy
moreover increases substantially with bond dimension in accordance with a
volume law. When we compare these two cases with intermediate cases selected
for the possibility of seeing a mobility edge, we find at the disorder strength
$\lambda=1.5$ evidence broadly
consistent with the claim that localization is present for lower energies
but not for energies towards the middle of the spectrum.
The case for delocalization is weaker at $\lambda=1.5$, but in both cases we
cannot make conclusive inferences from the data taken.

In addition to studying the same systems at larger bond dimension, we could
seek precise criteria for distinguishing states close to and far from true
eigenstates. In the absence of such a criterion, we are unable to say for
certain when the apparent saturation of entropy actually corresponds to having
found true localized states.  It would be further useful to establish a rigorous
theoretical relationship between the bond dimension and entropy for MPS
approximating extended excited states akin to the finite-entanglement scaling
relationship found for critical systems in \cite{critical-entanglement1}. We
leave this effort for future work. 

It may also be worthwhile in future work to modify the numerical techniques in
order to study bond-dimension scaling. For example, it may be useful to take a
candidate eigenstate from a lower bond dimension as an initial state (rather than
a random state) in order to see how robust that state is. It may similarly be
useful to track failure of convergence instead of simply designating a maximum
number of iterations and not distinguishing between ``convergence'' from the two
stopping criteria. Meanwhile, it may improve efficiency to allow bond dimension
to vary within a system (so that one may effectively save resources on ``weak''
bonds).

\noindent \textit{Note added:} After our initial submission of this work, there have been subsequent developments that indicate that 
well-studied localization transitions do not exist at the expected parameters:~\cite{abanin2021distinguishing, sels2021dynamical, sels2021markovian, morningstar2022avalanches, sierant2022challenges} in particular,
in the random-field Heisenberg model, a standard workhorse for studying MBL, the 
critical disorder strength once expected to be around $W = 3$ may instead 
be as high as $W_c\sim 20$~\cite{sels2021markovian, morningstar2022avalanches}.
Our results do not directly address these questions. It is more focused instead on developing
systematic matrix product methods to investigate aspects of localization transitions and,
in particular, the possible existence of a many-body mobility edge in the GAA model. 

\section*{Acknowledgements}
The authors would like to thank Jed Pixley and Bryan Clark for helpful
discussions. We additionally thank Rutgers University for their hospitality
during the workshop  ``Quasiperiodicity and Fractality in Quantum Statistical
Physics'', where part of this work was completed. S.G. was supported by the
National Science Foundation under Grant OMA-1936351. T.-C.W. and N.P. were
supported by the National Science Foundation under Grant PHY-1915165.
During revision of this work N.P. was 
supported by the National Science Foundation under Grant OMA-2120757.
The  authors would like to thank Stony Brook Research Computing and
Cyberinfrastructure, and the Institute for Advanced Computational Science at
Stony Brook University for access to the high-performance SeaWulf computing
system, which was made possible by the National Science Foundation under
Grant No. 1531492.

\bibliography{references.bib}

\input{app_include}

\end{document}

%% file: app_include.tex
\newcommand\shiftfigbox[2]{\raisebox{-#1}[\height-#1][0cm]{#2}}

\newlength\histwidthset
\setlength\histwidthset{.75\textwidth}

\clearpage
\appendix

\begin{widetext}
\begin{figure*}
  \newlength\mpsspacing     
  \newlength\mpsvert        
  \newlength\lbreak         
  \newlength\rbreak         
  \newlength\mpsunderset    
  \newlength\outset         
  \newlength\mpshwidth      
  \newlength\mpsheight      
  \newlength\mpohwidth      
  \newlength\mpohheight     
  \newlength\actionunderset 
  \newlength\mpsactwidth    
  \pgfmathsetlength\mpsspacing{.8cm}
  \pgfmathsetlength\mpsvert{.9cm}
  \def\nmps{8}
  \pgfmathtruncatemacro\nmmps{\nmps-1}
  \def\ihole{4}
  \pgfmathtruncatemacro\iihole{\ihole+1}
  \pgfmathtruncatemacro\iiihole{\ihole+2}
  \pgfmathtruncatemacro\imhole{\ihole-1}
  \pgfmathsetlength\lbreak{(\ihole-.45)*\mpsspacing}
  \pgfmathsetlength\rbreak{(\ihole+1.45)*\mpsspacing}
  \pgfmathsetlength\mpsunderset{.1\mpsvert}
  \pgfmathsetlength\outset{.05\mpsspacing}
  \pgfmathsetlength\mpshwidth{.30\mpsspacing}
  \pgfmathsetlength\mpsheight{.4\mpsvert}
  \pgfmathsetlength\mpohwidth{.25\mpsvert}
  \pgfmathsetlength\mpohheight{.35\mpsspacing}
  \pgfmathsetlength\actionunderset{.25\mpsvert}
  \pgfmathsetlength\mpsactwidth{\mpsspacing+\mpshwidth}
  \newlength\lbreakr                   
  \newlength\rbreakr
  \pgfmathsetmacro\eqdisp{.8+\nmps}     
  \pgfmathsetmacro\rhdisp{.8+\eqdisp}   
  \pgfmathsetlength\lbreakr{\rhdisp\mpsspacing+\lbreak}
  \pgfmathsetlength\rbreakr{\rhdisp\mpsspacing+\rbreak}
  \tikzstyle{phistyle}=[blue!50!cyan,draw=blue]
  \tikzstyle{psistyle}=[magenta!70,draw=blue!50!red]
  \tikzstyle{mpostyle}=[green!60,rounded corners,draw=green!70!black]
  \begin{tikzpicture}
    \foreach \i in {0,...,\nmps} {
      \node[coordinate] (ket\i) at (\i\mpsspacing,0) {};
      \node[coordinate] (kmpo\i) at (\i\mpsspacing,\mpsvert) {};
      \node[coordinate] (bmpo\i) at (\i\mpsspacing,2\mpsvert) {};
      \node[coordinate] (bra\i) at (\i\mpsspacing,3\mpsvert) {};
    }
    \draw (ket0) -- (\lbreak,0);
    \draw (\rbreak,0) -- (ket\nmps);
    \draw (bra0) -- (\lbreak,3\mpsvert);
    \draw (\rbreak,3\mpsvert) -- (bra\nmps);
    \draw (kmpo0) -- (kmpo\nmps);
    \draw (bmpo0) -- (bmpo\nmps);
    \foreach \a/\b in {0/\imhole,\iiihole/\nmps}
    \foreach \i in {\a,...,\b} {
      \draw (ket\i) -- (bra\i);
    }
    \draw (\ihole\mpsspacing,.6\mpsvert) -- (\ihole\mpsspacing,2.4\mpsvert);
    \draw (\iihole\mpsspacing,.6\mpsvert) -- (\iihole\mpsspacing,2.4\mpsvert);
    \filldraw[phistyle] (ket0){}++(-\outset,-\mpsunderset) --
      +(0,\mpsheight) -- +(\mpshwidth,0) -- cycle;
    \filldraw[phistyle] (bra0){}++(-\outset,\mpsunderset) --
      +(0,-\mpsheight) -- +(\mpshwidth,0) -- cycle;
    \filldraw[phistyle] (ket\nmps){}++(\outset,-\mpsunderset) --
      +(0,\mpsheight) -- +(-\mpshwidth,0) -- cycle;
    \filldraw[phistyle] (bra\nmps){}++(\outset,\mpsunderset) --
      +(0,-\mpsheight) -- +(-\mpshwidth,0) -- cycle;
    \foreach \a/\b in {1/\imhole,\iiihole/\nmmps}
    \foreach \i in {\a,...,\b} {
      \filldraw[phistyle] (ket\i){}++(0,-\mpsunderset) --
        +(-\mpshwidth,0) -- +(0,\mpsheight) -- +(\mpshwidth,0) -- cycle;
      \filldraw[phistyle] (bra\i){}++(0,\mpsunderset) --
        +(-\mpshwidth,0) -- +(0,-\mpsheight) -- +(\mpshwidth,0) -- cycle;
    }
    \pgfsetfillcolor{green!60}
    \pgfsetstrokecolor{green!70!black}
    \foreach \i in {0,...,\nmps} {
      \filldraw[mpostyle] (kmpo\i){}+(-\mpohwidth,-\mpohheight) --
        +(\mpohwidth,-\mpohheight) -- +(\mpohwidth,\mpohheight) --
        +(-\mpohwidth,\mpohheight) -- cycle;
      \filldraw[mpostyle] (bmpo\i){}+(-\mpohwidth,-\mpohheight) --
        +(\mpohwidth,-\mpohheight) -- +(\mpohwidth,\mpohheight) --
        +(-\mpohwidth,\mpohheight) -- cycle;
    }
    \pgfsetstrokecolor{black}
    \draw (bra0){}+(\lbreak,\actionunderset) -- +(\rbreak,\actionunderset);
    \draw (bra\ihole){} ++(0,\actionunderset) -- +(0,-.6\mpsvert);
    \draw (bra\iihole){}++(0,\actionunderset) -- +(0,-.6\mpsvert);
    \path (bra\ihole){} node[coordinate] (action0) at ++(0,\actionunderset) {};
    \filldraw[cyan!80!yellow,draw=cyan!30!blue] (action0){}+(-\mpshwidth,\mpsunderset)
      -- +(0,-\mpsheight) -- +(\mpsspacing,-\mpsheight)
      -- +(\mpsactwidth,\mpsunderset) -- cycle;
    \node[font=\Large] at (\eqdisp\mpsspacing,1.5\mpsvert) {$=$};
    \foreach \i [evaluate=\i as \x using {\i+\rhdisp}] in {0,...,\nmps} {
      \node[coordinate] (rket\i) at (\x\mpsspacing,.5\mpsvert) {};
      \node[coordinate] (rmpo\i) at (\x\mpsspacing,1.5\mpsvert) {};
      \node[coordinate] (rbra\i) at (\x\mpsspacing,2.5\mpsvert) {};
    }
    \draw (rbra0) -- (rbra\nmps);
    \draw (rket0) -- (\lbreakr,.5\mpsvert);
    \draw (\rbreakr,.5\mpsvert) -- (rket\nmps);
    \draw (rmpo0) -- (rmpo\nmps);
    \foreach \a/\b in {0/\imhole,\iiihole/\nmps}
    \foreach \i in {\a,...,\b} {
      \draw (rket\i) -- (rbra\i);
    }
    \draw (rbra\ihole) -- +(0,-1.4\mpsvert);
    \draw (rbra\iihole) -- +(0,-1.4\mpsvert);
    \filldraw[phistyle] (rket0){}++(-\outset,-\mpsunderset) --
      +(0,\mpsheight) -- +(\mpshwidth,0) -- cycle;
    \filldraw[psistyle] (rbra0){}++(-\outset,\mpsunderset) --
      +(0,-\mpsheight) -- +(\mpshwidth,0) -- cycle;
    \filldraw[phistyle] (rket\nmps){}++(\outset,-\mpsunderset) --
      +(0,\mpsheight) -- +(-\mpshwidth,0) -- cycle;
    \filldraw[psistyle] (rbra\nmps){}++(\outset,\mpsunderset) --
      +(0,-\mpsheight) -- +(-\mpshwidth,0) -- cycle;
    \foreach \i in {1,...,\nmmps} {
      \filldraw[psistyle] (rbra\i){}++(0,\mpsunderset) --
        +(-\mpshwidth,0) -- +(0,-\mpsheight) -- +(\mpshwidth,0) -- cycle;
    }
    \foreach \a/\b in {1/\imhole,\iiihole/\nmmps}
    \foreach \i in {\a,...,\b} {
      \filldraw[phistyle] (rket\i){}++(0,-\mpsunderset) --
        +(-\mpshwidth,0) -- +(0,\mpsheight) -- +(\mpshwidth,0) -- cycle;
    }
    \foreach \i in {0,...,\nmps} {
      \filldraw[mpostyle] (rmpo\i){}+(-\mpohwidth,-\mpohheight) --
        +(\mpohwidth,-\mpohheight) -- +(\mpohwidth,\mpohheight) --
        +(-\mpohwidth,\mpohheight) -- cycle;
    }

    \path[font=\large] (ket0){}
        node[rotate=90] at +(-.75cm,0) {$\bra{\psi_{i+1}}\ $}
        node[rotate=90] at +(-.75cm,\mpsvert) {$\;O^\dagger$}
        node[rotate=90] at +(-.75cm,2\mpsvert) {$O$}
        node[rotate=90] at +(-.75cm,3\mpsvert) {$\ \ket{\psi_{i+1}}$};

    \path[font=\large] (rket\nmps){}
        node[rotate=90] at +(.75cm,0) {$\bra{\psi_{i+1}}\ $}
        node[rotate=90] at +(.75cm,\mpsvert) {$O$}
        node[rotate=90] at +(.75cm,2\mpsvert) {$\ket{\psi_{i}}$};

  \end{tikzpicture}
  \caption{The tensor equation underlying the SIMPS algorithm developed in
  Yu, Pekker, \& Clark\cite{SIMPS} : for
  $O=H-E_0$, update $\ket{\psi_{i+1}}$ by solving for the light-blue
  tensor $T$ (and then decompose $T$ using SVD).
  The purpose of this is to optimize $\bra{\psi_{i+1}}O^\dagger\ket{\psi_i}$
  subject to the constraint $\bra{\psi_{i+1}}O^\dagger O\ket{\psi_{i+1}}=1$,
  but with a Lagrange multiplier, absorbed into $T$, that can be solved for with
  normalization.}
  \label{fig:simps}
\end{figure*}
\end{widetext}

\section{The SIMPS algorithm}
\label{app:algorithm}

In order to perform the SIMPS algorithm, as with other DMRG-based MPS
algorithm, we begin by expressing the Hamiltonian as an automaton-style
matrix product operator, formed in this case from the Jordan-Wigner transform
of the fermionic Hamiltonian, expressed in terms of operator-valued matrices
as
\begin{align*}
    O_n &= \left(\begin{array}{ccccc}
        \mathbbm{1} &0&0&0&0\\
        \sigma_-&0&0&0&0\\
        \sigma_+&0&0&0&0\\
        \frac{\sqrt{V}}{2} \sigma_z&0&0&0&0\\
        \frac{W_n}{2}\sigma_z-E_0\mathbbm{1}&\sigma_z\sigma_+&\sigma_-\sigma_z&\frac{\sqrt{V}}{2}\sigma_z&\mathbbm{1}\\
    \end{array}\right),\\
    W_n &= 2\lambda\frac{\cos(2\pi b n+\psi)}{1-\alpha\cos(2\pi b n+\phi)},
\end{align*}
with boundary vectors $v_L=(0,0,0,0,1)$ and $v_R=(1,0,0,0,0)$.

\begin{figure}[h!]
  \includegraphics[width=.45\textwidth]{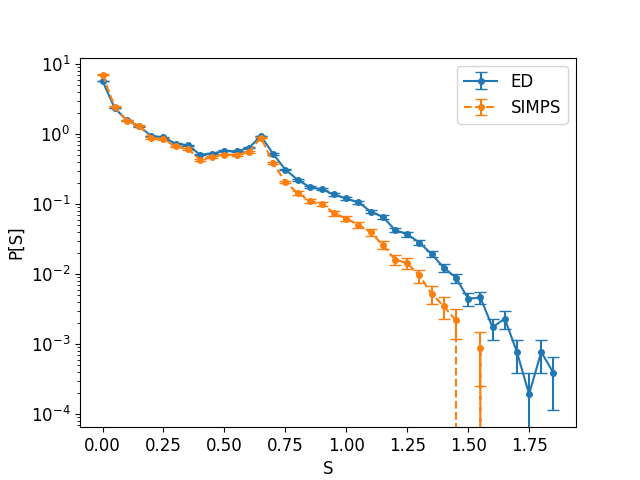}\\
  \caption{A replication of the test run by Yu, Pekker, and Clark to
  eliminate entropy bias of the SIMPS algorithm from consideration.
  The disordered Heisenberg model, with disorder strength $W=8$, is analyzed
  for 101 disorder realizations on a 10-site spin chain. Each of the
  $2^L$ eigenstates are extracted via exact diagonalization; then
  SIMPS, with bond dimension $\chi=12$,
  is used for $2^L$ equally-spaced target energies, with
  post-processing to remove duplicates (if $|\braket{\psi_i}{\psi_j}|>0.3$, we
  exclude the state with greater $\Delta E^2$). This produces 103,424
  states via exact diagonalization and 45,917 states via SIMPS. The latter
  is about three times as many states as in the original test, which explains
  the difference in noise in that case. As with the original visualization,
  SIMPS yields significantly fewer states at nearly all entropy ranges,
  the exceptions being the particularly small ones ($S\leq 0.2$) and the
  ``resonance'' peak at $S=\log 2$, with the difference in frequency being
  visibly greater at, e.g., $S\sim 1$ compared to $S\sim 0.5$.}
  \label{fig:bias-confirm}
\end{figure}

To find an eigenstate,
\begin{enumerate}
\item Start with an initial matrix-product ansatz $\ket{\psi_0}$
and a target energy $E_0$, incorporated into the MPO as above.
\item Given an iteration $\ket{\psi_i}$, optimize the next iteration
$\ket{\psi_{i+1}}$ as follows:
\item $\ket{\psi_{i+1}}$ may be initialized randomly, but we have, following a
suggestion by Clark\cite{clark-private},
initialized it with $\ket{\psi_i}$
\item Site-by-site, optimize $\ket{\psi_{i+1}}$ to satisfy
$(H-E_0)\ket{\psi_{i+1}}=\psi_i$: that is, apply the shifted and inverted
Hamiltonian.
\item To do so, we represent this equation as the maximization of
$\bra{\psi_{i+1}}(H-E_0)\ket{\psi_i}$,
subject to the constraint
$\bra{\psi_{i+1}}(H-E_0)^2\ket{\psi_{i+1}}=\|\psi_i\|^2$
which will be uniquely satisfied by $(H-E_0)^{-1}\psi_i$.
\item This is done, site-by-site, by solving the diagrammatic equation in
Fig.~\ref{fig:simps} for individual tensors. We note that we find it preferable
to update two sites at once (i.e. the tensor being optimized consists of the
contraction of tensors at two sites), especially when enforcing
charge/fermion-number conservation, in order to speed up convergence. The
resulting two-site tensor is then split via SVD to update the MPS.
\item This may be repeated until $\ket{\psi_{i+1}}$ has converged;
alternatively, when initializing $\psi_{i+1}$ with $\psi_i$, very few sweeps
(optimizing the tensors at each site) may be conducted per iteration, as the
goal of convergence is the eigenvalue equation $\psi\propto(H-E_0)\psi$, which
should be more accurate after each sweep.
\item Repeat until the energy has converged, or until a maximum number
of iterations has been reached.
\end{enumerate}

\begin{figure}
  \includegraphics[width=.45\textwidth]{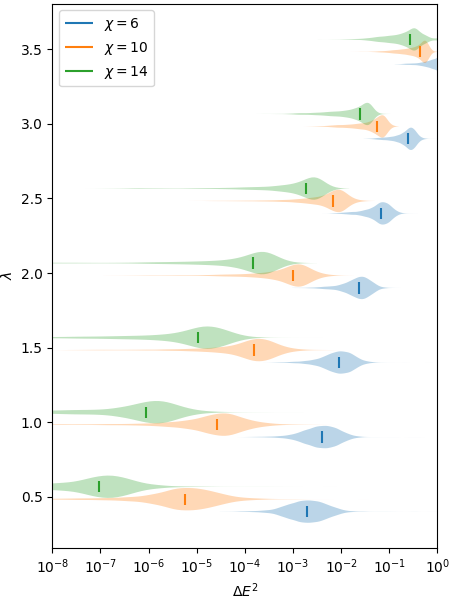}
  \caption{The distribution of energy
  errors at several (smaller) bond dimensions, given system size
  $L=128$, for tests sampling the full spectra of systems with
  $\lambda \in \{0.5,1.0,1.5,2.0,2.5,3.0,3.5\}$}
  \label{fig:bwL128}
\end{figure}

\begin{figure}
  \includegraphics[width=.45\textwidth]{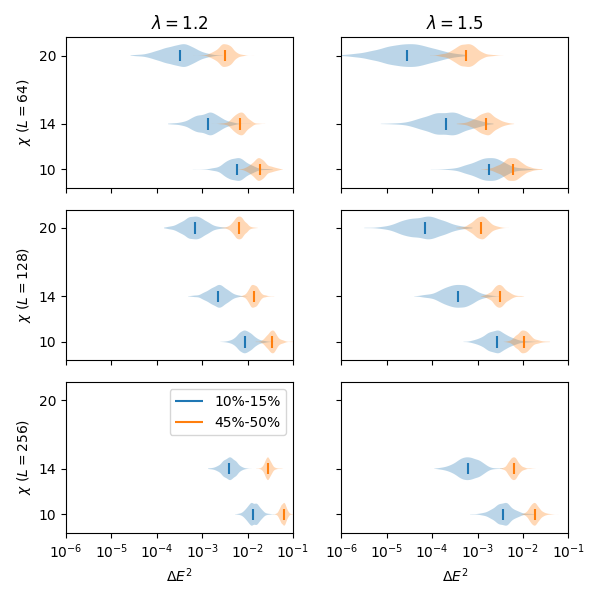}\\
  \caption{A violin plot displaying the distribution of energy errors, in
  the systems under consideration within the main text but for a greater
  selection of system sizes: for $\chi\in\{10,14,20\}$, the systems
  are analyzed at $L=128$, and for $\chi\in\{10,14\}$, the systems are analyzed
  at $L=256$. Note that the the $L=64$ data is contained within
  Fig.~\ref{fig:energy-violin} of the main text. This plot demonstrates that,
  with these parameters, system size does not affect the quality of states
  within the sizes considered.}
  \label{fig:error-allL}
\end{figure}

In the original work of Yu, Pekker, and Clark \cite{SIMPS}, the authors claim
that SIMPS is ``sampling states at a given entanglement with the same
frequency as ED and hence there is no systemic bias.''
This would be quite remarkable,
given the general expectation that entropy may diverge approaching a transition
- in particular, for any bond dimension $\chi$ there should be
\textit{truly localized} states with entanglement entropy at some cut
in excess of the maximum $\log\chi$. (In fact, in a good approximation of
a physical state, it may be expected that the entropy ceiling should be
even \text{less} than that absolute maximum, as, for example, was shown for
infinite MPS approximations for ground states of critical spin chains by
Pollmann et al. \cite{critical-entanglement1}.) Although they acknowledge
a ``failure of SIMPS to find high-quality eigenstates in [the] near-ergodic and
ergodic regime'', they do not explain why there should be a hard boundary
between regimes near to and far from ergodicity. Moreover, in the data they
provide as evidence for this claim (Fig.~S2), the divergence of the proportion
of SIMPS states from that of ED states at higher entanglement
entropy seems apparent (if small), and likely statistically
significant. To confirm statistical significance, we replicate the test they
use to produce these data as faithfully as possible, yielding data
that clearly replicates the major features of this figure, particularly a
divergence between sampling rates at higher entropies, in
Fig.~\ref{fig:bias-confirm}. 

In addition to favoring true low-entropy eigenstates, we have noted that the
SIMPS algorithm will produce ``false'' eigenstates when no low-entropy
eigenstates are available, as is evidenced by the fact that the algorithm
produces any states at all within the presumed ergodic regime. To attempt to
constrain the false eigenstates we observe, we may try to approximate a
worst-case scenario by supposing that there exist $n$ consecutive eigenstates,
of some separation $s$: that is, taking the crudest possible approximation, the
energies take the form $E_k=E_0+ks$. Then the energy variance would be
\begin{equation}
\begin{split}
\Delta E^2 &\equiv \langle H^2\rangle - \langle H\rangle^2 =\langle (H-E_0)^2\rangle - \langle H-E_0\rangle^2\\
&= \sum_{k=0}^{n-1}\frac{s^2k^2}{n} - \left(\sum_{k=0}^{n-1}\frac{s k}{n}\right)^2= \frac{n(n-1)}{12}s^2.
\end{split}
\label{eq:energy-error-calc}
\end{equation}
We do not presuppose the order of magnitude of $n$, as the possible
entropy reduction from such a combination is highly dependent on the nature
of the eigenstates themselves. We may, however, presume that the worst-case
energy spacing $s$ is of the order $2^{-L}$, such that

\begin{equation}
\Delta E^2 \sim \frac{n^2}{12}\frac{L^2}{2^{2L}}.
\label{eq:energy-error-approx}
\end{equation}

For the system sizes under consideration this is well below machine precision;
it will likely be necessary to include further assumptions, e.g. from the
random matrix theory formulation of ETH, in order to find a reasonable lower
bound. 

\begin{figure}
  \includegraphics[width=.38\textwidth]{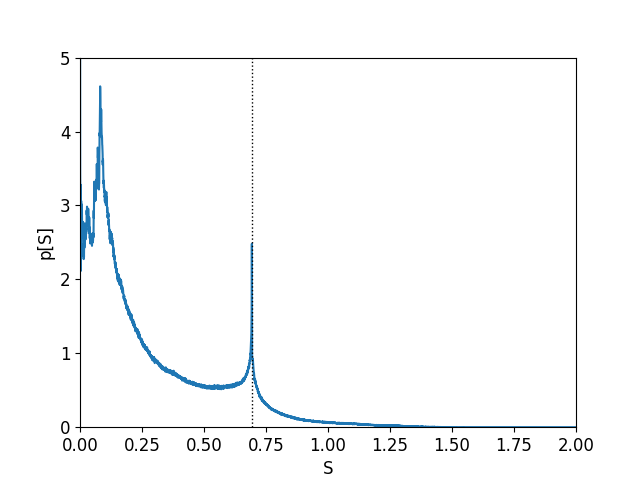}
  \caption{A histogram of all single-cut entropies, for all states, of
  12-site systems with $\lambda=2$, given 1920 ``disorder realizations''
  (i.e. phases $\phi$). This demonstrates a very clear resonance peak at
  $S=\ln 2$, marked by the dotted line.}
  \label{fig:SED}
\end{figure}

\begin{figure*}
\centering
  \subfigure[$\ell=1$]{\shiftfigbox{.3cm}{\includegraphics[width=\histwidthset]{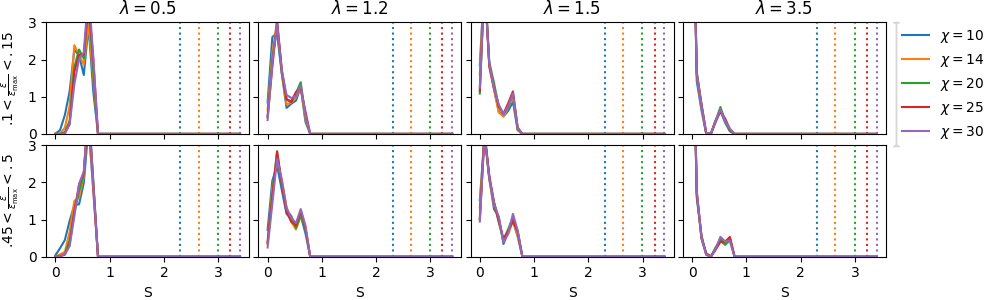}}}
  \subfigure[$\ell=2$]{\shiftfigbox{.3cm}{\includegraphics[width=\histwidthset]{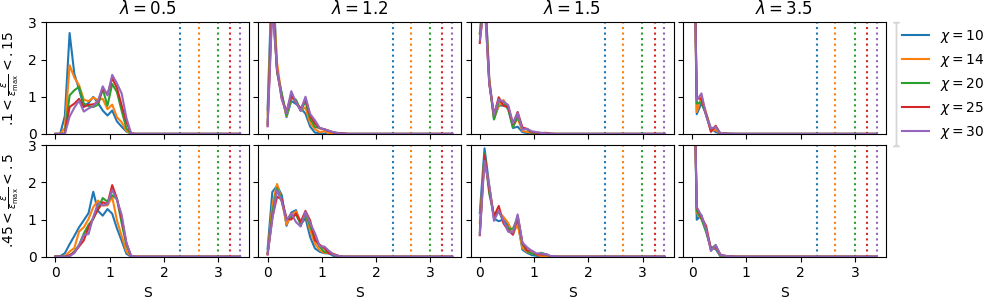}}}
  \subfigure[$\ell=3$]{\shiftfigbox{.3cm}{\includegraphics[width=\histwidthset]{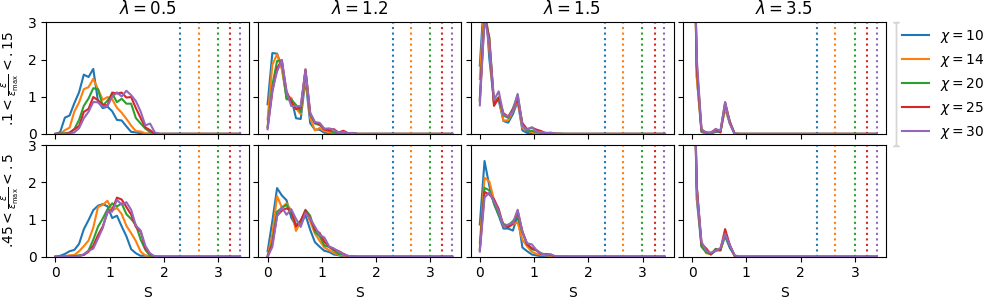}}}
  \subfigure[$\ell=4$]{\shiftfigbox{.3cm}{\includegraphics[width=\histwidthset]{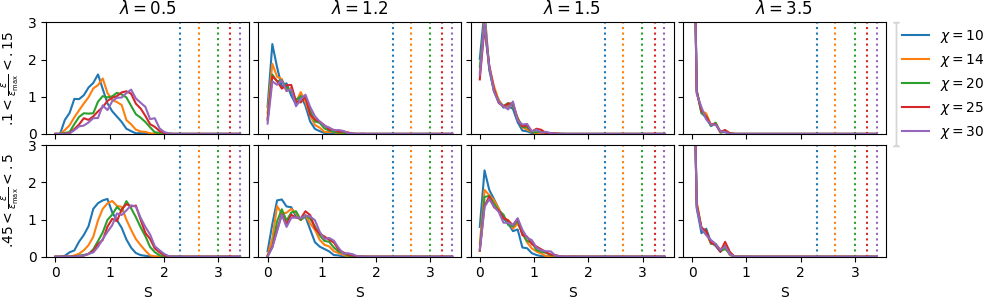}}}
  \subfigure[$\ell=5$]{\shiftfigbox{.3cm}{\includegraphics[width=\histwidthset]{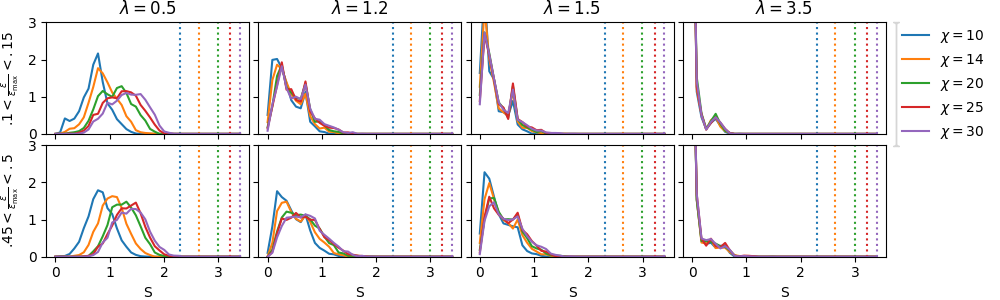}}}
  \caption{Histograms of single-cut entanglement entropy by cut position,
  corresponding to the data in Fig.~\ref{fig:trend-by-cut} of the main
  text}
  \label{fig:ShistbycutA}
\end{figure*}

\begin{figure*}
\centering
  \subfigure[$\ell=6$]{\shiftfigbox{.3cm}{\includegraphics[width=\histwidthset]{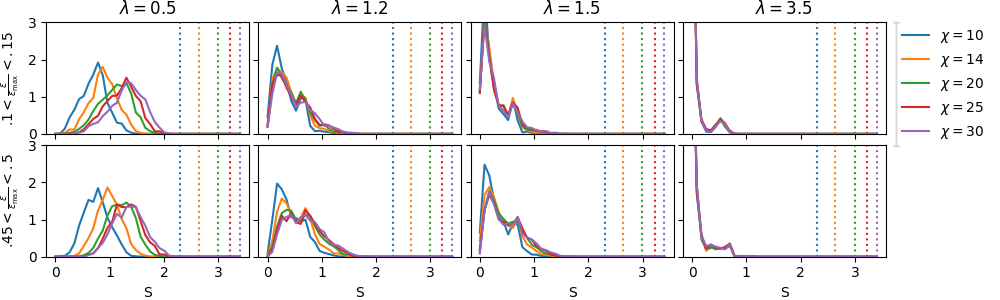}}}
  \subfigure[$\ell=7$]{\shiftfigbox{.3cm}{\includegraphics[width=\histwidthset]{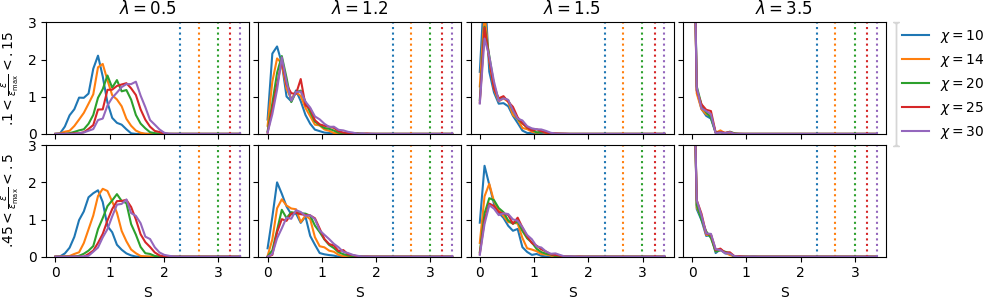}}}
  \subfigure[$\ell=8$]{\shiftfigbox{.3cm}{\includegraphics[width=\histwidthset]{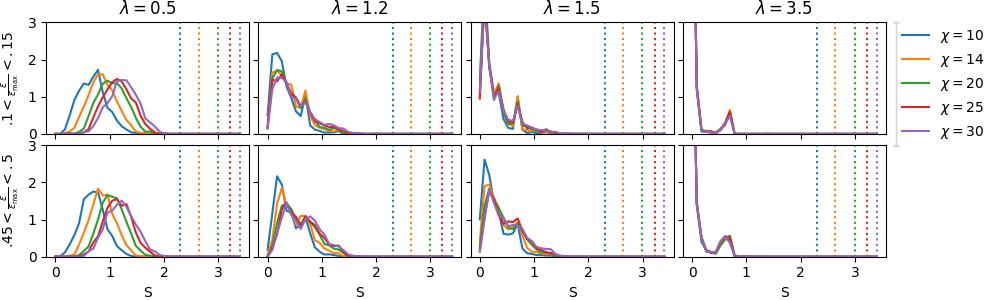}}}
  \subfigure[$\ell=9$]{\shiftfigbox{.3cm}{\includegraphics[width=\histwidthset]{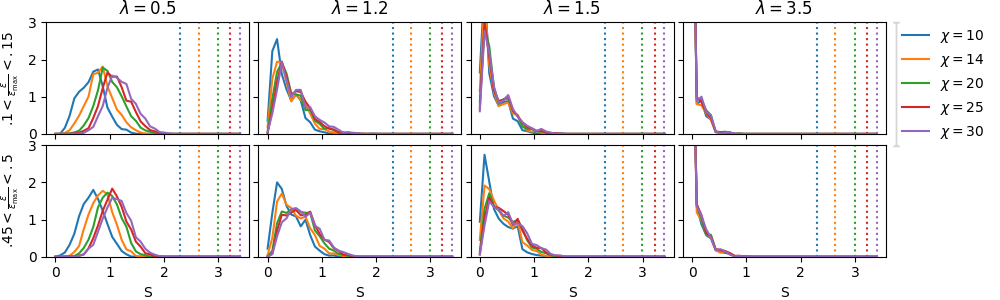}}}
  \subfigure[half-cut]{\shiftfigbox{.3cm}{\includegraphics[width=\histwidthset]{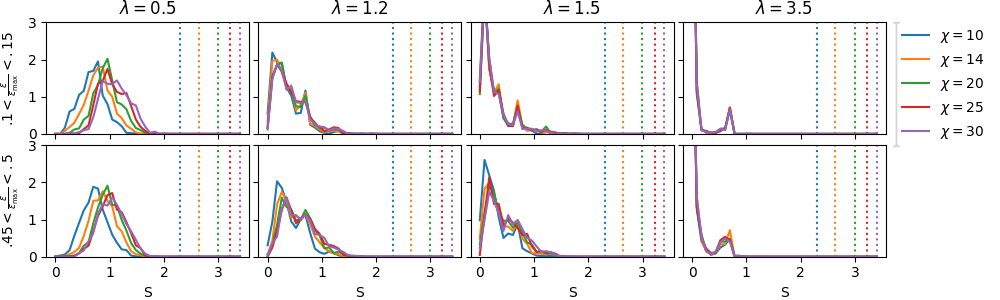}}}
  \caption{Histograms of single-cut entanglement entropy by cut position,
  corresponding to the data in Fig.~\ref{fig:trend-by-cutB} of the main
  text, as well as (e) histograms of half-cut entanglement entropy
  (i.e. $\ell=32$ for $L=64$)}
  \label{fig:ShistbycutB}
\end{figure*}

\begin{figure*}
\centering
  \subfigure[$\lambda=1.2$, lower energy range]{\shiftfigbox{.3cm}{\includegraphics[width=\histwidthset]{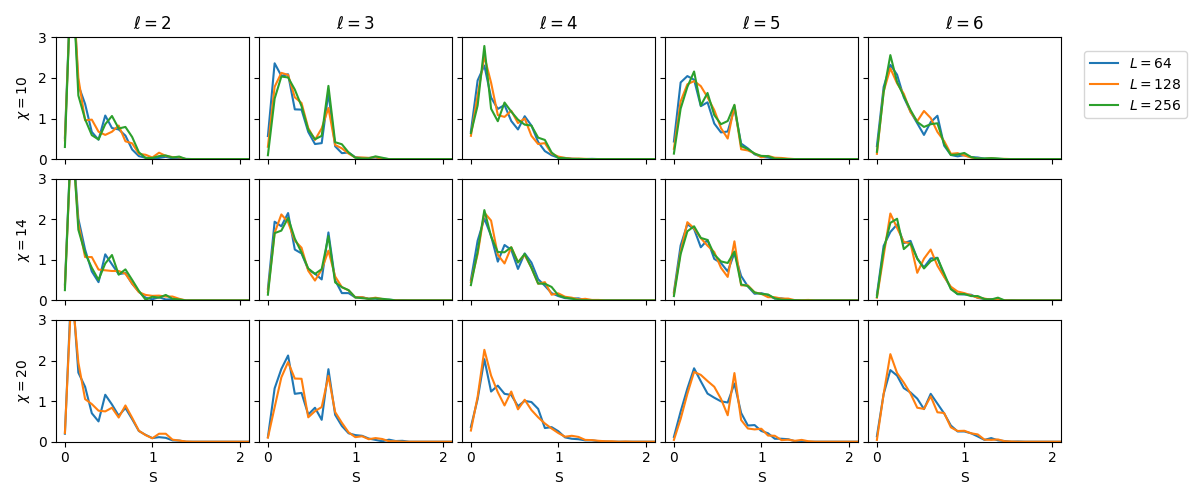}}}
  \subfigure[$\lambda=1.2$, higher energy range]{\shiftfigbox{.3cm}{\includegraphics[width=\histwidthset]{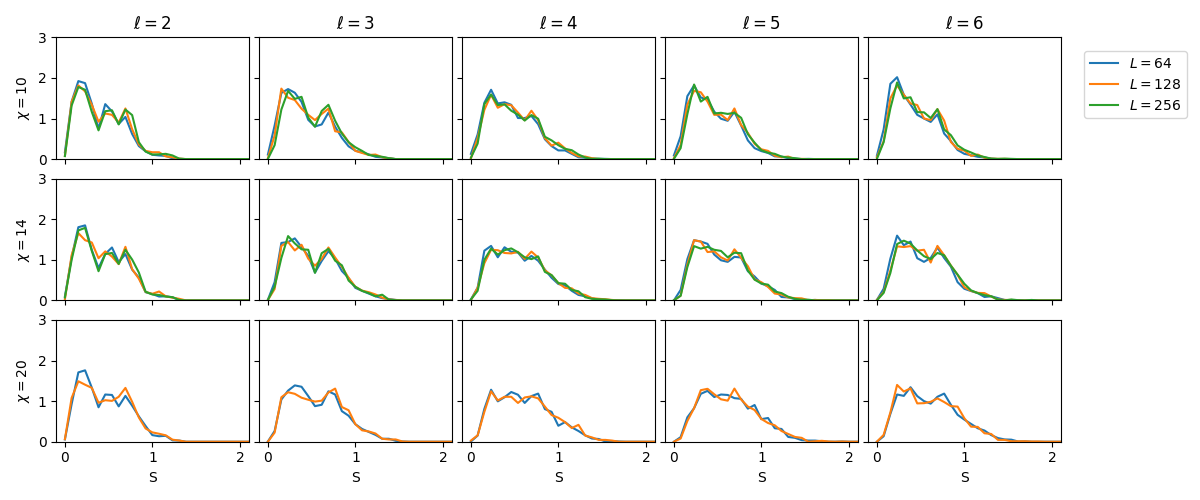}}}
  \subfigure[$\lambda=1.5$, lower energy range]{\shiftfigbox{.3cm}{\includegraphics[width=\histwidthset]{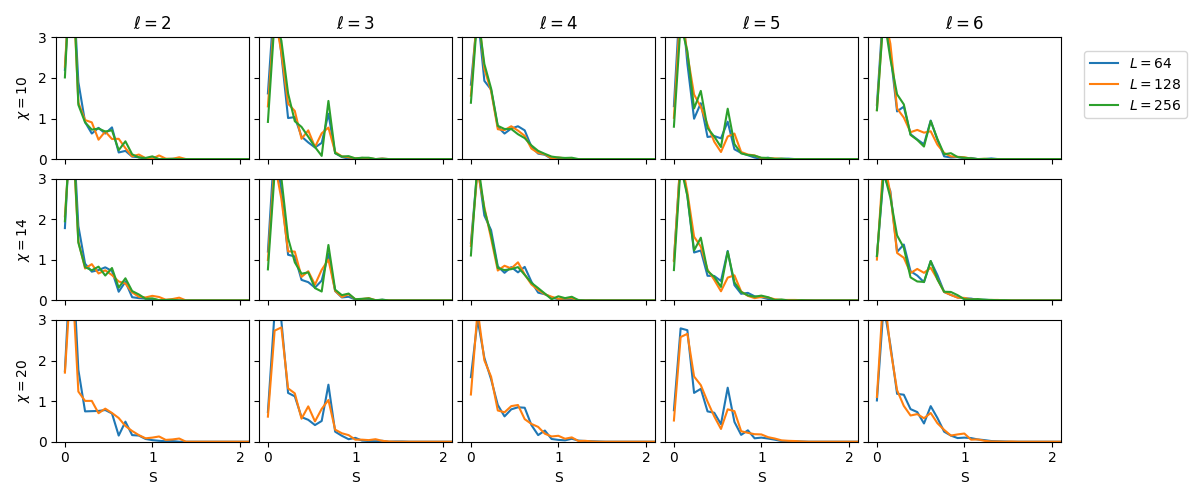}}}
  \subfigure[$\lambda=1.5$, higher energy range]{\shiftfigbox{.3cm}{\includegraphics[width=\histwidthset]{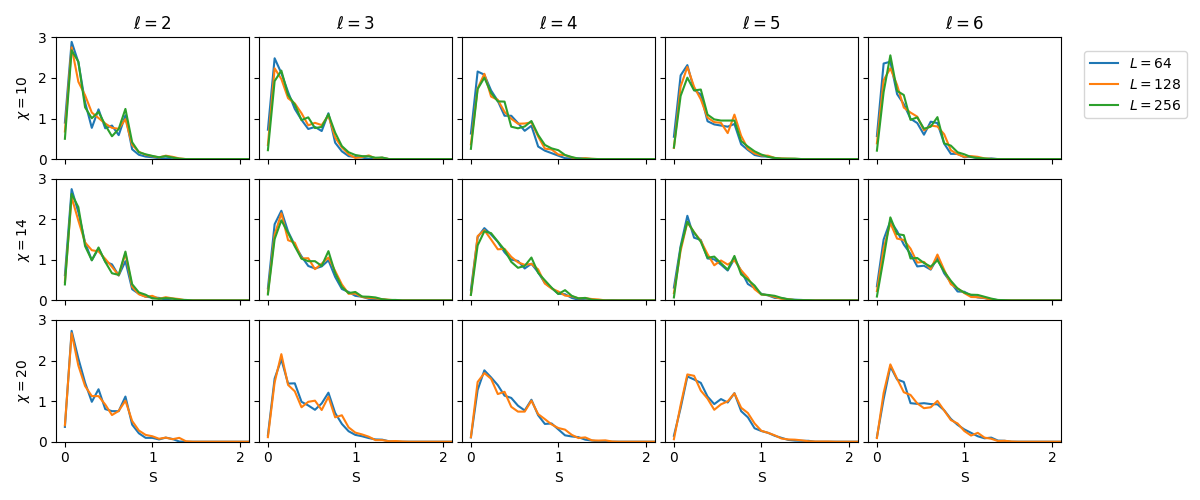}}}
  \caption{Histograms of single-cut entanglement entropy in different regimes,
  plotted for different cut locations (from $\ell=2$ to $\ell=6$) and
  bond dimensions and compared across lengths.
  }
  \label{fig:Shistbycutlong}
\end{figure*}

\begin{figure}
  \includegraphics[width=.45\textwidth]{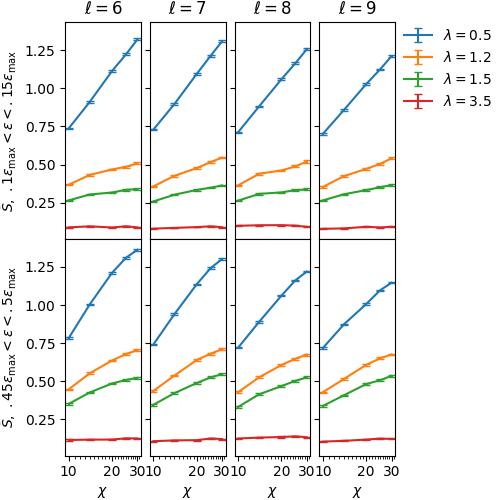}
  \caption{Average single-cut entanglement entropy, together with the
  standard error of the mean, as a function of the bond dimension $\chi$ for
  several larger cuts (as in Fig.~\ref{fig:trend-by-cut} of the main
  text but with the subsequent four cuts).}
  \label{fig:trend-by-cutB}
\end{figure}

\section{Entropy oscillations}
\label{app:osc}
A characteristic effect seen in the localized regime (and even at higher bond dimensions) is the
oscillation of single-cut entanglement entropy as a function of cut location; this
effect is most clearly seen in Fig.~\ref{fig:entropy-cut3.5} but is also visible in
Fig.~\ref{fig:entropy-intermediate} at higher bond dimensions. The more detailed histograms in
Figures~\ref{fig:ShistbycutA} and~\ref{fig:ShistbycutB} show peaks in the entanglement entropy
which seem less prominent across even-numbered cuts and in more thermalized
conditions. We confirm in Fig.~\ref{fig:SED} that these peaks are not spurious
by analyzing a collection of systems at size $L=12$ with exact diagonalization;
this additionally lets us conclude with high confidence that the typical
location of the peaks is at $S=\ln 2$. We may understand these entropy peaks
as indicating the presence of a dimer: taking a pair of qubits in the
``single-occupancy'' subspace $\operatorname{span}\{\ket{01},\ket{10}\}$,
uniformly distributed over the Bloch sphere, define a random variable
$\mathrm{S}_\text{dimer}$ to be the entanglement entropy between them.
Then the various histograms under consideration are, for the most part, qualitatively
consistent with drawing the entropy from $\mathrm{S}_1+\mathrm{S}_\text{dimer}$ with probability
$p_\text{dimer}$ and $\mathrm{S}_0$ with probability $1-p_\text{dimer}$, given
a pair of well-behaved, unimodal random variables $\mathrm{S}_0$ and $\mathrm{S}_1$.

It is worth noting as well that such entropy
oscillations are often seen in
models of spinless fermion chains with tight-binding couplings or
(equivalently, under the Jordan-Wigner transformation)
Heisenberg-like antiferromagnetic spin chains when open boundary conditions break 
translation symmetry \cite{affleck2009entanglement,laflorencie2006boundary,wang2004boundary}.
In particular, Laflorencie et al.~\cite{laflorencie2006boundary} have identified this
effect as a dimerization process universal in ground states of models with a Luttinger
liquid description (such as the critical XXZ chain), determining that the alternating parts
of the energy and entropy, $E_A$ and $S_A$ respectively, should have proportional universal
contributions which may be expressed as
\begin{equation*}
    S_A(\ell,L) \propto E_A(\ell,L) \sim \left[\frac{L}{\pi} \sin\left(\tfrac{\pi\ell}{L}\right)\right]^{-K}
\end{equation*}
for the $\ell$th cut of a length-$L$ chain in a system with Luttinger parameter $K$.

\section{Additional datasets}
\label{app:datasets}

In addition to the dataset described and referenced in the main text, we have
two additional datasets that we will reference on occasion in these
Supplemental Materials. As in the main text, each uses the Hamiltonian defined
by \eqref{eq:GAA-ham} and \eqref{eq:GAA-interaction}, with $\alpha=0.3$,
$t=V=1$, and $b=\frac{2}{1+\sqrt{5}}$, and with protection of $U(1)$ symmetry
to restrict to half-filling.
\subsection{Exploratory trials}
We began by testing a wide range of disorder strengths across the spectrum.
For disorder strengths $\lambda$ in 0.5, 1.0, 1.5, 2.0,
  2.5, 3.0, and 3.5 (that is, the half-integers between 0 and 4), we took 24
  disorder realizations (that is, values of $\phi$ in \eqref{eq:GAA-ham}). With
  bond dimension $\chi=10$, we analyzed systems of size $L=16,32,64,128$; we
  additionally applied bond dimensions $\chi=6$ and $\chi=14$ to systems of size
  $L=128$. In each case, we selected 400 target energies that encompass the full
  energy spectrum (noting that this means we would see, and reject, a number of
  copies of the states with lowest and highest energy). We have used this
  dataset to produce Figures~\ref{fig:bwchi10},~\ref{fig:bwL128},
  and~\ref{fig:Shistold}. In the latter, we also have a subset of those
  conditions, namely $\chi=10$ and $\chi=14$ for $L=128$, with
  $\lambda=1.2$.

\subsection{Comparison of system sizes}
\label{app:sys-size}
By examining the system at smaller length scales, we have been able to
conclude that at least the most dramatic finite-size effects do not persist
into the system size where the main trials were conducted.
We have additionally taken the parameters for the ``intermediate regime''
under primary consideration, $\lambda=1.2$ and $\lambda=1.5$ with
energy density $0.1\varepsilon_\text{max}< \varepsilon< 0.15\varepsilon_\text{max}$
and $0.45\varepsilon_\text{max}< \varepsilon < 0.5\varepsilon_\text{max}$, and
extracted candidate eigenstates as in the main trials with the larger system sizes 
$L=128$ (for $\chi=10,\ 14,\ \&\ 20$) and $L=256$ (for $\chi=10\ \& \ 20$).
In Fig.~\ref{fig:error-allL}, we examine the energy error from these trials and
find that it does not vary substantially when we increase system size. We have
also been able to use the single-cut entanglement entropy to observe the boundary
effects, seeing in Figures~\ref{fig:entropy-cut0.5} and~\ref{fig:entropy-intermediate}
that even the most persistent boundary effects do not appear to persist far enough
into the bulk to make a significant difference in the cases being considered. In
Fig.~\ref{fig:Shistbycutlong} we take a closer look by comparing entropy histograms at
various cuts across system sizes, finding that there is little consistent
variation as we increase system size (and that what variation there is, is not
monotonic, as the $L=256$ graphs tend to seem closer to the $L=64$ graphs than to the
$L=128$ ones.)

\section{Review of diagnostic metrics}
In the main text we have relied primarily on two metrics,
energy error and single-cut entanglement entropy, to evaluate the
goodness and behavior of candidate eigenstates. Here we review these
metrics, including more detailed plots summarizing entanglement-entropy
distributions, and then discuss several additional metrics not utilized in
the main text.
\subsection{Energy error}
\label{app:energy-error}

We have presented our primary results on the energy error in
Fig.~\ref{fig:energy-violin}; we show additional results for larger system
sizes in Fig.~\ref{fig:error-allL}. We here note briefly how it is calculated.
In particular, it emerges fairly easily from SIMPS calculation: contracting
the transfer matrices used to obtain the LHS of Fig.~\ref{fig:simps}
gives $\langle(H-E_0)^2\rangle$, and $\langle H-E_0\rangle$ (which does
\textit{not} come pre-calculated) is easily computed through the simpler
contraction of the RHS of the same (replacing $\ket{\psi_i}$ with
$\ket{\psi_{i+1}}$.)

In Figs.~\ref{fig:bwchi10} and~\ref{fig:bwL128}, we introduce an additional
set of simulations, in which we sampled the entire spectrum for a substantial
number of ``disorder'' strengths $\lambda$, with bond dimensions among
$\{6,10,14\}$ and lengths (for $\chi=10$; otherwise $L=128$) in
$\{16,32,64,128\}$, to examine how our results scale with bond dimension and
with system size.

\subsection{Single-cut entanglement entropy}
\label{app:entropy}

\begin{figure}
  \includegraphics[width=.45\textwidth]{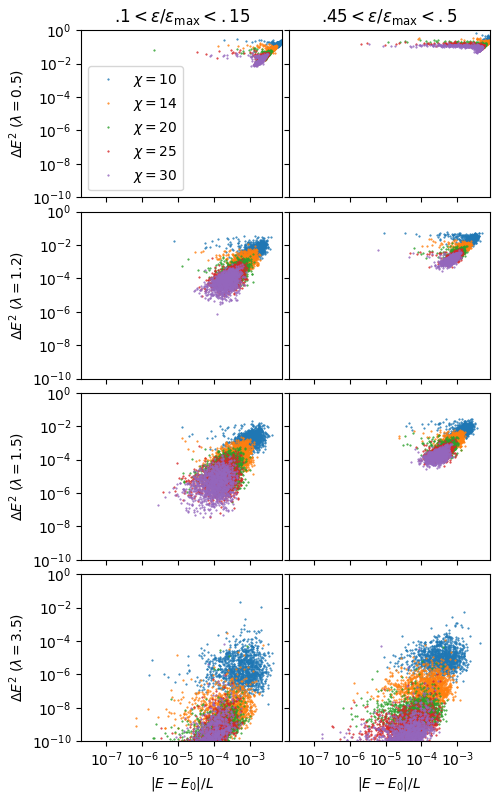}
  \caption{Scatter plot of energy error versus energy wandering, to attempt
  to determine whether a relation exists between quality of states \& how
  far from the target energy the algorithm must ``wander'' to find it}
  \label{fig:E0dE}
\end{figure}

\begin{figure*}
  \subfigure[width 3]{\includegraphics[width=.80\textwidth]{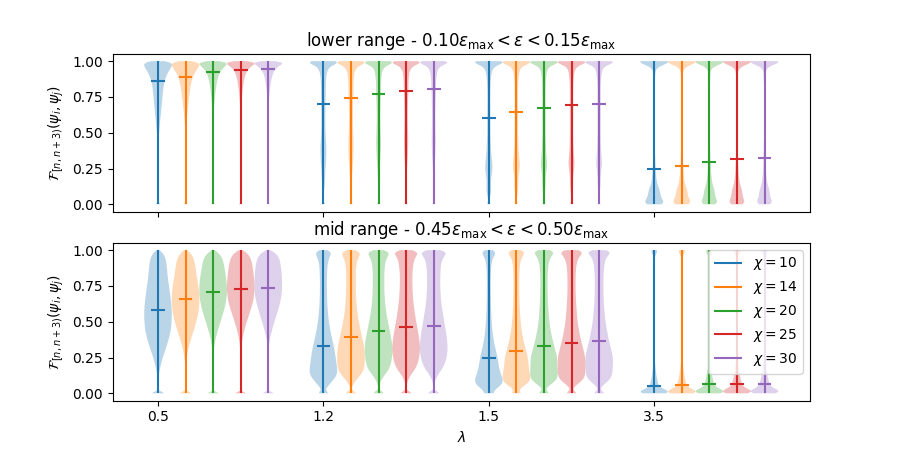}}
  \subfigure[width 5]{\includegraphics[width=.80\textwidth]{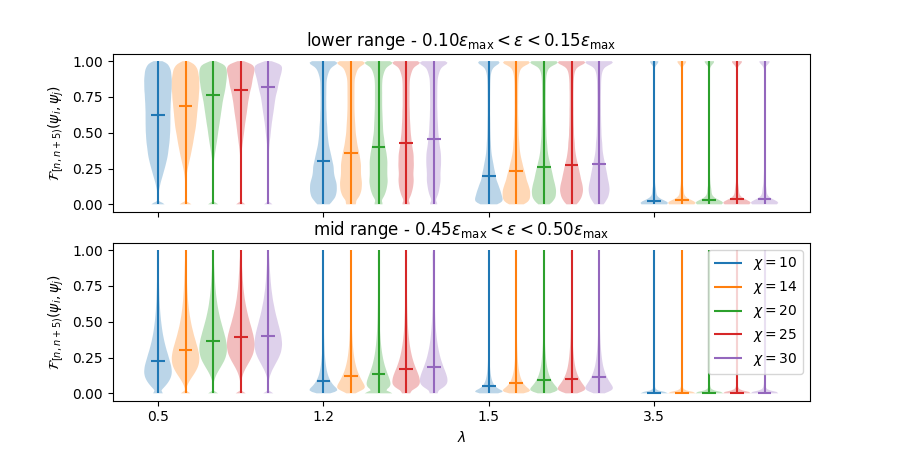}}
  \subfigure[width 8]{\includegraphics[width=.75\textwidth]{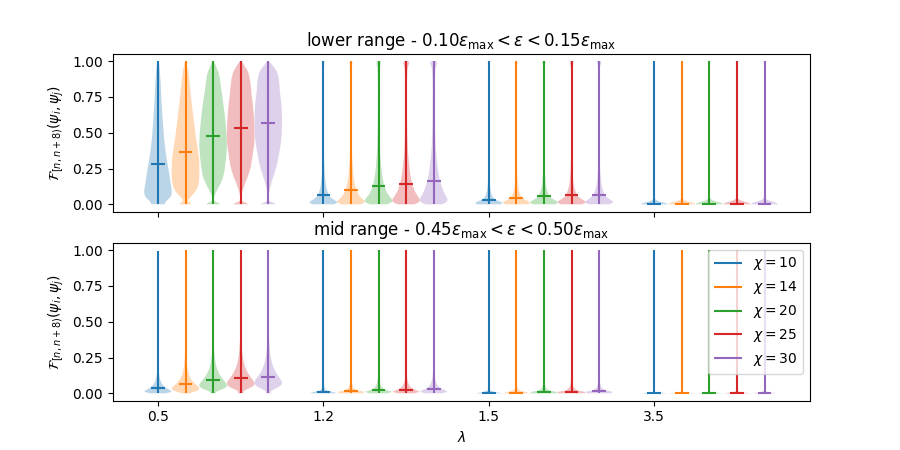}}
  \caption{Violin plots of Uhlmann fidelities among systems with various
  disorder strengths and at different energy ranges, for segments of width
  (a) 3, (b) 5, and (c) 8. All fidelities calculated are between states with
  the same disorder strength \textit{and disorder sample}, bond dimension,
  and energy range. All segments of the given length within the 64-site
  chain are considered.}
  \label{fig:uhlmann}
\end{figure*}

\begin{figure*}
  \includegraphics[width=.7\textwidth]{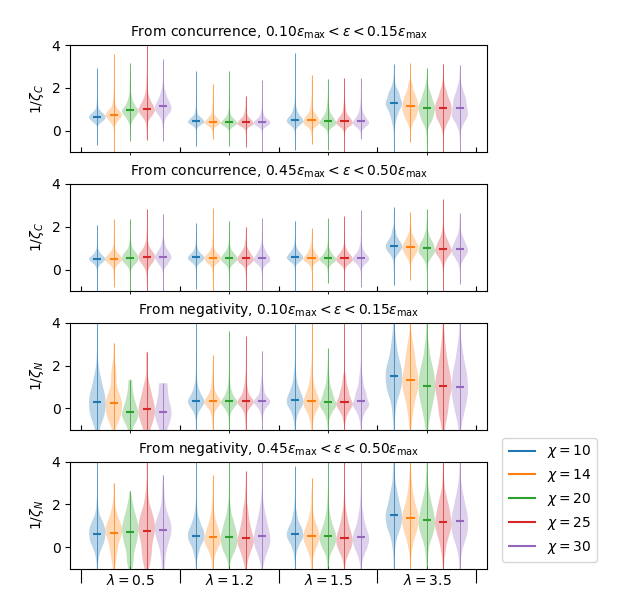}
  \caption{Violin plots of (inverse) localization lengths among systems with
  various disorder strengths and at different energy ranges, taking a weighted
  average of the (inverse) lengths for each state analyzed}
  \label{fig:loclenwavg}
\end{figure*}

\begin{figure*}
    \includegraphics[width=\textwidth]{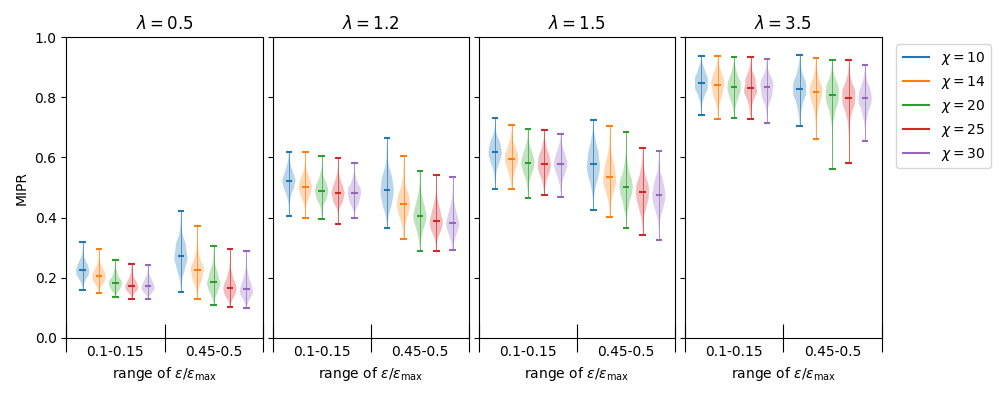}
    \caption{Violin plots of the many-body inverse participation ratio
    for states in the main datasets}
    \label{fig:mipr-short}
\end{figure*}
\begin{figure}
    \includegraphics[width=.5\textwidth]{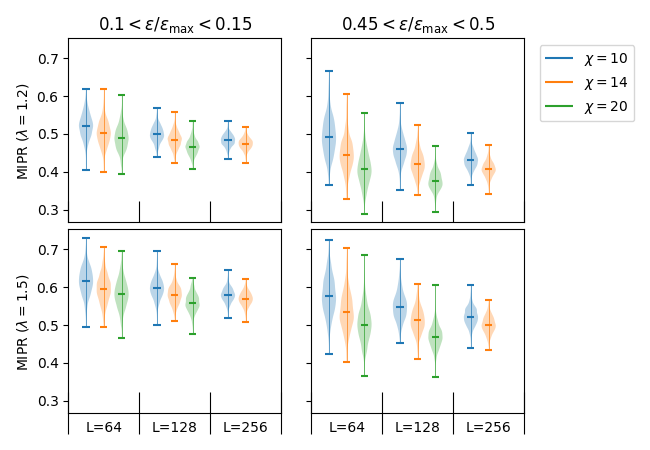}
    \caption{Violin plots comparing many-body inverse participation ratio across system sizes}
    \label{fig:mipr-long}
\end{figure}

By keeping the MPS in (bi)canonical form, we are able to extract entanglement
entropies directly from the Schmidt coefficients which are stored as part of
the ansatz. We have explored how average entropies, at given distances from the
boundary, scale with the bond dimension $\chi$ in Figs.~\ref{fig:trend-by-cut},
\ref{fig:entropy-cut3.5},~\ref{fig:entropy-cut0.5},
and~\ref{fig:entropy-intermediate} of the main text. In particular, in
Fig.~\ref{fig:trend-by-cut}, we examined the entropy scaling at several
specific cuts; we repeat this further into the bulk in
Fig.~\ref{fig:trend-by-cutB}, and then go deeper by examining the corresponding
entropy distributions in each case in Figures~\ref{fig:ShistbycutA}
and~\ref{fig:ShistbycutB}, respectively. Similar entropy distributions are examined in
Fig.~\ref{fig:Shistbycutlong}, there comparing entropy histograms at different system
sizes to investigate whether boundary effects show system-size dependence at the
primary length $L=64$ considered. In Fig.~\ref{fig:Shistold}, we take a
different approach and examine the entropy distribution for all cuts at various
$\lambda$, $L$, and $\chi$.
One feature that is clearly visible in many of these histograms is a
peak at $S=\ln 2$, corresponding to dimers or two-site resonances.
We confirm that this is not a numerical artifact using exact diagonalization in Fig.~\ref{fig:SED}.

\subsection{Energy wandering}
\label{app:wandering}

Another quantity we may use to diagnose the goodness of states is the so-called
``energy wandering'', the difference between the energy of a state and the
target energy used to obtain it. The idea behind using this is to determine
whether or not approximate eigenstates of adequate quality are sufficiently
common. In Fig.~\ref{fig:E0dE} we compare the distribution of values of
$\|E-E_0\|$ with that of $\Delta E^2$.
\subsection{Uhlmann fidelities}
\label{app:uhlmann}

Inspired by, and using methods based on, \cite{hauru-uhlmann}, we
compute Uhlmann fidelities: if the reduced density matrix of an eigenstate
$\psi_i$ on a segment $A$ is $\rho_{i,A}=\op{tr}_A\ket{\psi_i}\bra{\psi_i}$,
then the Uhlmann fidelity between $\psi_i$ and $\psi_j$ on $A$ is
\begin{equation}
\mathcal{F}=\op{tr}\sqrt{\sqrt{\rho_{i,A}}\rho_{j,A}\sqrt{\rho_{i,A}}}.
\label{eq:uhlmann}
\end{equation}
Ideally,
\begin{itemize}
\item In the localized case, the distribution of these quantities will be
determined by so-called ``l-bits'': if $\psi_i$ and $\psi_j$ differ on an
l-bit whose support is within $A$, then $\mathcal{F}=0$; if they agree on
all l-bits mostly supported within $A$, then $\mathcal{F}\sim 1$; and
intermediate values will only occur when there are l-bits on which
$\psi_i$ and $\psi_j$ differ that have significant support both inside and
outside of $A$.
\item In the fully ergodic regime, where $\rho_{i,A}$ should be fully determined
by the energy (as $\op{tr}_A\exp(-\beta(E_i) H)$, with $\beta(E_i)$ the
inverse-temperature corresponding to $E_i$), we expect a less discrete 
distribution of $\mathcal{F}$, with values continuously dependent on the
energy difference and stochastically dependent on the choice of region $A$.
\end{itemize}

In Fig.~\ref{fig:uhlmann}, we examine the distribution of Uhlmann fidelities
in various systems considered, for various sizes of region $A$. We see that
the typical behavior, in the localized case, is a bimodal distribution with
one narrow peak at 0 corresponding to cases differing on l-bits supported in $A$
and another narrow peak at 1 corresponding to cases agreeing on l-bits overlapping $A$,
with the former shrinking and the latter growing both as the size of $A$
increases (so that $\mathcal{F}=1$ would require agreement on more l-bits)
and as we move to the higher-energy band, which we expect to have larger
localization length.
In the delocalized case, meanwhile, we see a broad unimodal distribution
whose peak, in addition to lowering as the size of $A$ and the energy of the
band increase, raises as the bond dimension increases, suggesting an increase
in similarity as the accuracy improves. (It is not truly unimodal however;
a small peak at $\mathcal{F}=0$ which narrows with increasing bond dimension
suggests that the pseudo-eigenstates obtained in this case do sometimes have
features which resemble l-bits.

In analyzing the $\lambda=1.2$ and $\lambda=1.5$ cases, we find that
\begin{itemize}
  \item For the most part, the distribution in the higher-energy band is closer
  to a unimodal distribution like the one seen in the delocalized case; in (a) a
  second peak close to $\mathcal{F}=1$ in the width-3 distribution exists but
  grows less distinct with increasing bond dimension
  \item The distribution in the lower-energy band is more consistently
  bimodal, although with lower maxima at nonzero fidelity.
  \item In particular, as bond dimension increases the distributions in the
  higher-energy band appear to converge towards a unimodal distribution; this seems to
  be the case, though it is less clear, for the lower-energy band when
  $\lambda=1.2$. However, for the lower-energy band with $\lambda=1.5$,
  we see apparent convergence toward a bimodal distribution in (a) and (b).
\end{itemize}

\subsection{Localization lengths}
\label{app:loc-lengths}

We follow \cite{entanglement-measures1,entanglement-measures2} in using two
measures of entanglement between two qubits, negativity and concurrence, to
attempt to estimate the localization length, which should diverge approaching a
localization transition or mobility edge. In particular, given concurrence
values $C_{i,j}$ or negativity values $N_{i,j}$ between two sites of a given
state, we fit the nontrivial values to (see eqs. 21 and 22 of
\cite{entanglement-measures2})
\begin{equation}
\begin{split}
C_{i,i\pm n} &= k^{C\pm}_i\exp(-n/\zeta^{C\pm}_i)\\
N_{i,i\pm n} &= k^{N\pm}_i\exp(-n/\zeta^{N\pm}_i).
\end{split}
\end{equation}

In Fig.~\ref{fig:loclenwavg}, we take, for each state, an average
of all these $1/\zeta^{C\pm}_i$ and, separately, $1/\zeta^{N\pm}_i$, weighted by
$\frac{1}{\sigma_{1/\zeta}}$. While this confirms some basic expectations --
the localization lengths of eigenstates with $\lambda=3.5$ tend to be much
smaller, and for both types of entanglement the lengths are greater in the
middle band than in the lower band -- at other times the results are unexpected
or even self-contradictory, for example, when the typical localization length
appears to decrease with bond dimension and when it is lower for $\lambda=1.2$
and $\lambda=1.5$ than for $\lambda=0.5$. We must therefore conclude that we
will not be able to perform much meaningful analysis on this data.

\subsection{Inverse participation ratio}
\label{app:mipr}
In analysis performed following the initial submission of this work,
we apply the many-body generalization of the inverse participation ratio \eqref{eq:ipr},
which has been defined \cite{soumya-mipr,vu-fermionic-mbl}, for a system with
$N$ fermionic orbitals of which $N_f=\nu N$ are occupied,
\begin{equation}
    \mathcal{I} = \frac{1}{1-\nu}\left(\frac{1}{N_f}\sum_{i}\langle \hat{n}_i\rangle^2 - \nu\right)
\end{equation}
where $\hat{n}_i$ is the number operator for the $i$th orbital. This ensures that
this many-body inverse participation ratio (MIPR),
$\mathcal{I}$, interpolates between $\mathcal{I}=0$ in the ideal thermalized case where
all sites have equal filling, $\langle\hat{n}_i\rangle=\nu$, and $\mathcal{I}=1$
in the ideal localized case when filling is deterministic, $\langle\hat{n}_i\rangle=1$ with
probability $\nu$ and $\langle\hat{n}_i\rangle=0$ with probability $1-\nu$.

Here each site hosts one orbital, $N=L$, and half-filling or $\nu=1/2$ makes
\[ \mathcal{I} = -1+\frac{4}{L}\sum_i\langle\hat{n}_i\rangle^2. \]

In Fig.~\ref{fig:mipr-short} we examine the distribution of the MIPR in states obtained for
the main dataset. In the localized benchmark ($\lambda=3.5$) we find that the MIPR
is large and does not vary substantially with bond dimension, whereas in the
thermalized benchmark ($\lambda=0.5$) we find the MIPR to be small and consistently decreasing
with bond dimension. For the mobility-edge candidate regime, we find an intermediate MIPR that
is roughly constant with bond dimension for the lower-energy regime and decreasing with
bond dimension for the higher-energy regime, suggesting that both $\lambda=1.2$ and
$\lambda=1.5$ remain compatible with a many-body mobility edge.

In Fig.~\ref{fig:mipr-long} we use the MIPR to compare the main dataset with
states obtained in larger systems (i.e. those analyzed in
Fig.~\ref{fig:error-allL}), finding that
the distributions tighten but otherwise remain qualitatively similar.